\documentclass[superscriptaddress,amsmath,amssymb,nofootinbib,eqsecnum,a4paper,secnumarabic,11pt]{revtex4}   
\usepackage[pdftex]{graphicx}
\usepackage{hyperref}     
\usepackage{url} 
\usepackage{slashed,color}
\usepackage[utf8]{inputenc} 

\begin{document}
\begin{flushright}
DESY 19-003
\end{flushright}

\title{Current status and future prospects of the singlet-doublet dark matter model with $CP$ violation}
\author{Tomohiro Abe}
\affiliation{
  Institute for Advanced Research, Nagoya University,
  Furo-cho Chikusa-ku, Nagoya, Aichi, 464-8602 Japan
}
\affiliation{
  Kobayashi-Maskawa Institute for the Origin of Particles and the
  Universe, Nagoya University,
  Furo-cho Chikusa-ku, Nagoya, Aichi, 464-8602 Japan
}

\author{Ryosuke Sato}
\affiliation{Deutsches Elektronen-Synchrotron (DESY), Notkestra\ss e 85, D-22607 Hamburg, Germany}

\begin{abstract}
We discuss the singlet-doublet fermion dark matter model with $CP$ violation.
In this model, 
the $CP$ violation generates a pseudoscalar interaction of dark matter with the standard model Higgs boson.
Thanks to the pseudoscalar interaction, the model can evade the strong constraint from the
dark matter direct detection experiments while keeping the success of the thermal relic scenario.
The $CP$ violation also predicts signals in dark matter indirect detection experiments
and electric dipole moments (EDM) that can be as large as the current upper bound.
We investigate the constraints and prospects of the direct detection experiments, 
the indirect detection experiment, and the electron EDM.
We also investigate the stability of the Higgs potential.
Combining these observables, we find that heavy dark matter is disfavored. 
We also find it is possible to probe the Higgs funnel region by the combination of the
direct detection experiments and the measurements of the electron EDM 
if experiments for the electron EDM reach to ${\cal O}(10^{-32})~e$ cm in future.
\end{abstract}

\maketitle

\section{Introduction}
The thermal relic scenario is an attractive scenario to explain the origin of the dark matter (DM).
This scenario does not rely on the initial condition of our Universe, and it has many indications on model building of DM.
See for a review Ref.~\cite{1703.07364}.
In this scenario, 
the energy density of the DM in the Universe is determined by 
the annihilation of the DM particles into the standard model (SM) particles.
Therefore, this scenario requires DM-SM interactions, and we can expect nonzero DM-nucleon scattering cross section.
There are experiments that aim to detect the DM directly through this scattering process, such as
the Xenon1T~\cite{1805.12562}
and PandaX-II experiments~\cite{1708.06917}.
However, signals of the DM-nucleon scattering have not been detected yet.
This null result gives severe constraints on the DM models,
and it becomes important to investigate models which can avoid such constraints.

The simplest model which can avoid the constraints from the direct detection experiments is
an effective theory that is constructed by introducing a gauge singlet fermion DM \cite{Kanemura:2010sh, 1203.2064, 1205.3169, 1309.3561, Matsumoto:2014rxa, 1512.06458, Matsumoto:2016hbs, 1808.10465}.\footnote{A similar but nonminimal approach by using a two-Higgs doublet model
with $Z_3$ was discussed in Ref.~\cite{1712.09873}.}
The DM in this setup does not couple to the SM field via renormalizable interactions.
The DM-Higgs interactions are realized by dimension-5 operators, $\bar{\psi} \psi H^{\dagger} H$ and $\bar{\psi} i \gamma^5 \psi H^{\dagger} H$.
If the coefficient of the $CP$-conserving operator $\bar{\psi} \psi H^{\dagger} H$ is negligible,
then the $CP$-violating operator $\bar{\psi} i \gamma^5 \psi H^{\dagger} H$ controls both of the amount of the relic abundance and the DM-nucleon scattering cross section.
The $CP$-violating operator generates a pseudoscalar interaction between DM and the SM Higgs boson, $\bar{\psi} i\gamma^5 \psi h$.
Since the nonrelativistic DM-nucleon scattering cross section with this interaction is highly suppressed by the
relative velocity between the DM and the nucleon, 
we can avoid the constraints from the direct detection experiments while keeping the amount of the relic abundance \cite{1203.2064}.
The interaction term is nonrenormalizable, and a cutoff scale accompanies it. 
The value of the cutoff scale to be required to obtain the correct relic abundance is ${\cal O}(1)$~TeV.
This result motivates us to consider UV completions of the effective theory as physics at TeV scale.

\begin{figure*}[t]
\centering
\includegraphics[width=0.48\hsize]{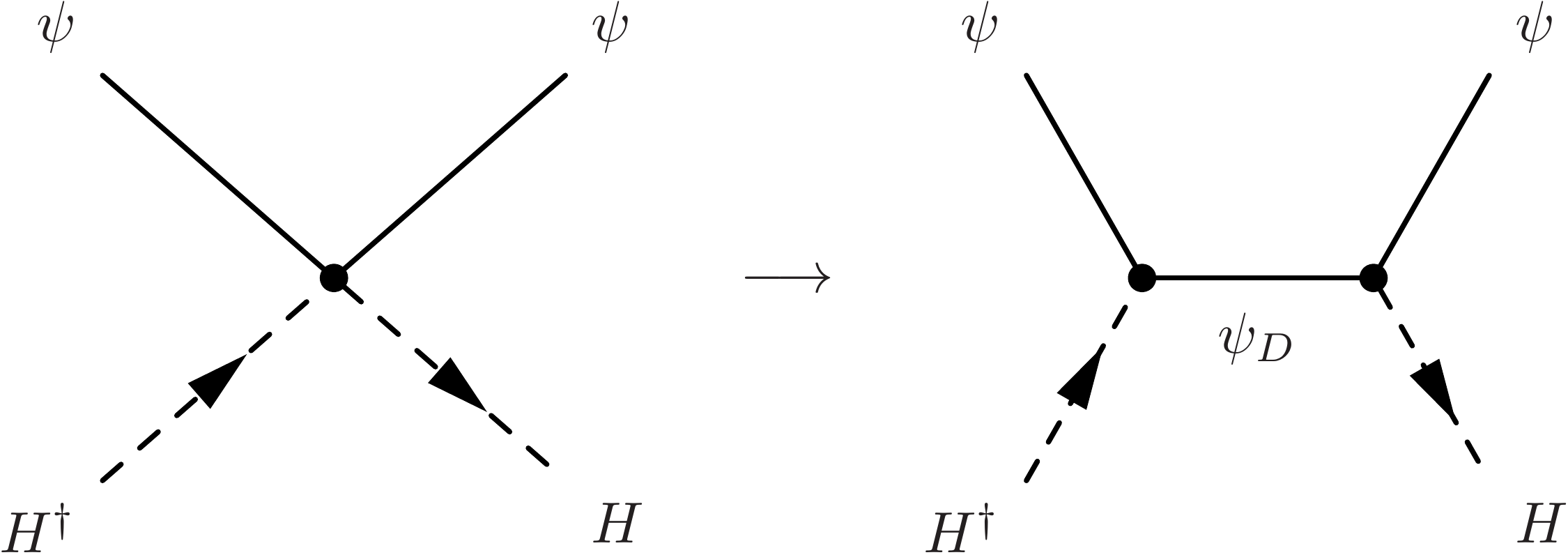} 
\caption{
The schematic view of the UV completion from the effective theory to the singlet-doublet model.
}\label{fig:UVcompletion}
\end{figure*}

The singlet-doublet DM model~\cite{hep-ph/0510064, 0705.4493, 0706.0918} is 
one of the UV completion of the above effective theory.
The model contains a gauge singlet Majorana fermion and an SU(2) doublet Dirac fermion
with hypercharge 1/2. This setup can be regarded as a generalization of the Bino-Higgsino system 
in the minimial supersymmetric standard model~\cite{hep-ph/0004043, hep-ph/0406144, hep-ph/0601041, 1510.05378}.
In the large Dirac mass limit, the model is reduced into 
the effective theory as schematically shown in Fig.~\ref{fig:UVcompletion}.
The model contains a $CP$ phase, and thus the pseudoscalar interaction exists.
Although this model has been widely studied~\cite{1109.2604,1211.4873,1311.5896,1505.03867,1506.04149,1509.05323,1510.05378,1512.02472,1602.04788,1603.07387,1611.02186, 1611.05048,1701.02737,1701.05869, 1707.03094,1708.01614},
its DM phenomenology with $CP$ violation has not been studied sufficiently.
See Refs.~\cite{1411.1335, 1702.07236} for previous studies.
The DM annihilation processes with the $s$-channel exchange of the Higgs boson with the pseudoscalar interaction are $s$ wave,
and its cross section at low temperature is not suppressed by powers of velocity.
As a result, the annihilation cross section of the DM in the current Universe can be sizable
enough to generate some signals in cosmic rays such as $\gamma$ rays.
Dwarf spheroidal galaxies (dSphs) are regarded as good targets to observe such $\gamma$-ray emissions because of
the less astrophysical uncertainty.
Using the Fermi-LAT data~\cite{Fermi-LAT:2016uux} that measures the $\gamma$-ray flux from dSphs, 
we can investigate the $CP$ violation in the singlet-doublet DM model.

Another consequence of the $CP$ violation is electric dipole moments (EDMs).
The predicted value of the electron EDM (eEDM) by the SM is $|d_e| \leq 10^{-38}~e$ cm~\cite{hep-ph/0504231, Pospelov:1991zt}. 
This is much smaller than the current upper bound on the eEDM by the ACME experiment~\cite{Andreev:2018ayy},
$|d_e| \leq 1.1 \times 10^{-29}~e$ cm (90\% C.L.).
It is known that the singlet-doublet model predicts the eEDM that can be as
large as the current upper bound~\cite{hep-ph/0510064, 0705.4493, 1411.1335, 1702.07236}.
Since the constraints from the direct detection experiments and the measurements of the eEDM are 
strong enough to probe the model, it is worth studying the correlation among the eEDM and other
 observables, such as the DM-nucleon scattering cross section,
to see if the model provides a viable DM candidate.

The stability of the Higgs potential also gives a bound on the singlet-doublet model.
Higgs couplings to fermions can give negative contributions to the beta function 
of the Higgs quartic coupling and make the Higgs potential unstable~\cite{1812.11303}.
While the pseudoscalar interaction of the DM with the Higgs boson
is essential in the singlet-doublet model to avoid the strong constraint from the direct detection experiments, 
it can make the Higgs potential unstable.
Therefore, imposing the stability on the Higgs potential gives another constraint on the singlet-doublet model.

The rest of this paper is organized as follows.
In Sec.~\ref{sec:eft}, we briefly review the current status of the minimal fermionic dark matter model.
We show the constraints on the $CP$-conserving and $CP$-violating DM-Higgs boson interactions.
It is found that the $CP$-conserving interaction is severely constrained from the DM direct detection,
and the $CP$ violation implies that a UV complete model appears around ${\cal O}(1)$ TeV scale.
In Sec.~\ref{sec:f12}, we discuss the singlet-doublet fermion model.
We focus on the constraint from the direct detection, indirect detection,
the stability of the Higgs potential, and the eEDM. 
The Higgs invisible decay is also discussed for $m_\text{DM} < m_h/2$.
It is shown that the heavy dark matter is disfavored by the combination of the constraints.
We also show that most of the region of the parameter space, including the Higgs funnel region,
can be probed by the combination of the direct detection experiments and the measurements of the eEDM in future. 
In Sec.~\ref{sec:summary}, we present our conclusion.

\section{Effective theory for fermionic DM models}\label{sec:eft}
\label{sec:eff}
In this section, we discuss a $Z_2$-odd gauge singlet Majorana fermion ($\psi$) as a DM candidate.
If we do not introduce any other new particle,
the DM cannot couple to the SM particles at the renormalizable level
and we need higher dimensional operators. 
We can write dimension-five operators with the Higgs boson.
\begin{align}
 {\cal L}
=&
 {\cal L}_{\text{SM}}
 + \frac{1}{2} \bar{\psi} \left( i \gamma^{\mu} \partial_\mu  - m_{\psi} \right) \psi
 + \frac{c_S}{2} \bar{\psi} \psi \left( H^\dagger H - \frac{v^2}{2}\right)
 + \frac{c_P}{2} \bar{\psi} i \gamma_5 \psi \left( H^\dagger H - \frac{v^2}{2}\right)
,
\label{eq:eff_lag}
\end{align}
where $H$ is the SM Higgs field.
There are two operators, $\bar{\psi}\psi H^\dagger H$ and
$\bar{\psi} i \gamma_5 \psi H^\dagger H$.
The former is a $CP$-conserving operator, and the latter violates the $CP$ invariance.
The $CP$-conserving interaction has been studied in Ref.~\cite{Kanemura:2010sh},
and the $CP$-violating operator has been studied in Refs.~\cite{1203.2064, 1205.3169, 1309.3561, 1512.06458,1808.10465}. 

These two operators have different properties in the DM physics.
In nonrelativistic DM-nucleon scattering processes, only the $CP$-conserving operator is relevant.
Thus, the spin-independent cross section is proportional to $c_S^2$,
\begin{align}
 \sigma_{\text{SI}}
~=~
\frac{1}{\pi}
\frac{f_N^2 c_S^2}{m_h^4}
\frac{m_N^4 m_{\text{DM}}^2}{(m_N + m_{\text{DM}})^2}
~\simeq~
\left(\frac{c_S}{\text{TeV}^{-1}}\right)^2
3.22 \times 10^{-44} ~\text{cm}^2
,
\end{align}
where $m_N$ is the nucleon mass.
Here we have used the following numbers.
\begin{align}
 m_N =& 0.938 \text{ GeV},\\
 f_N =& \frac{2}{9} + \frac{7}{9} \sum_q f_q,\\
 f_u = 0.0110,\ \
 f_d =& 0.0273,\ \
 f_s = 0.0447.
\end{align}
where $f_q$ values are taken from
\texttt{micrOMEGAs}~\cite{1407.6129}.

On the other hand, both operators contribute to the DM annihilation processes,
which are important to determine the dark matter relic density. 
The annihilation cross section is the $p$-wave ($s$-wave) process with the $CP$-conserving ($CP$-violating) operator.

\begin{figure*}[t]
\centering
\includegraphics[width=0.48\hsize]{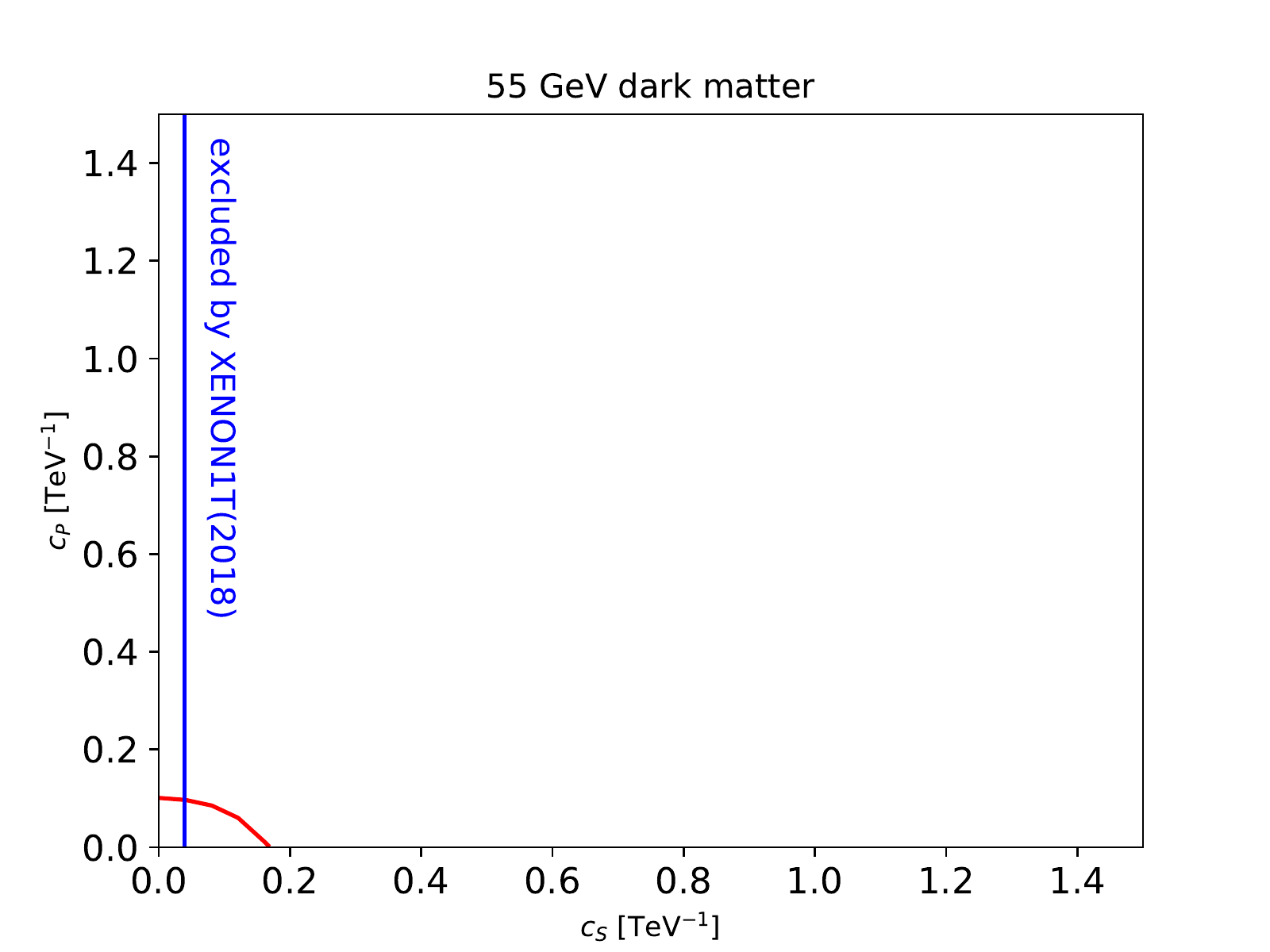} 
\includegraphics[width=0.48\hsize]{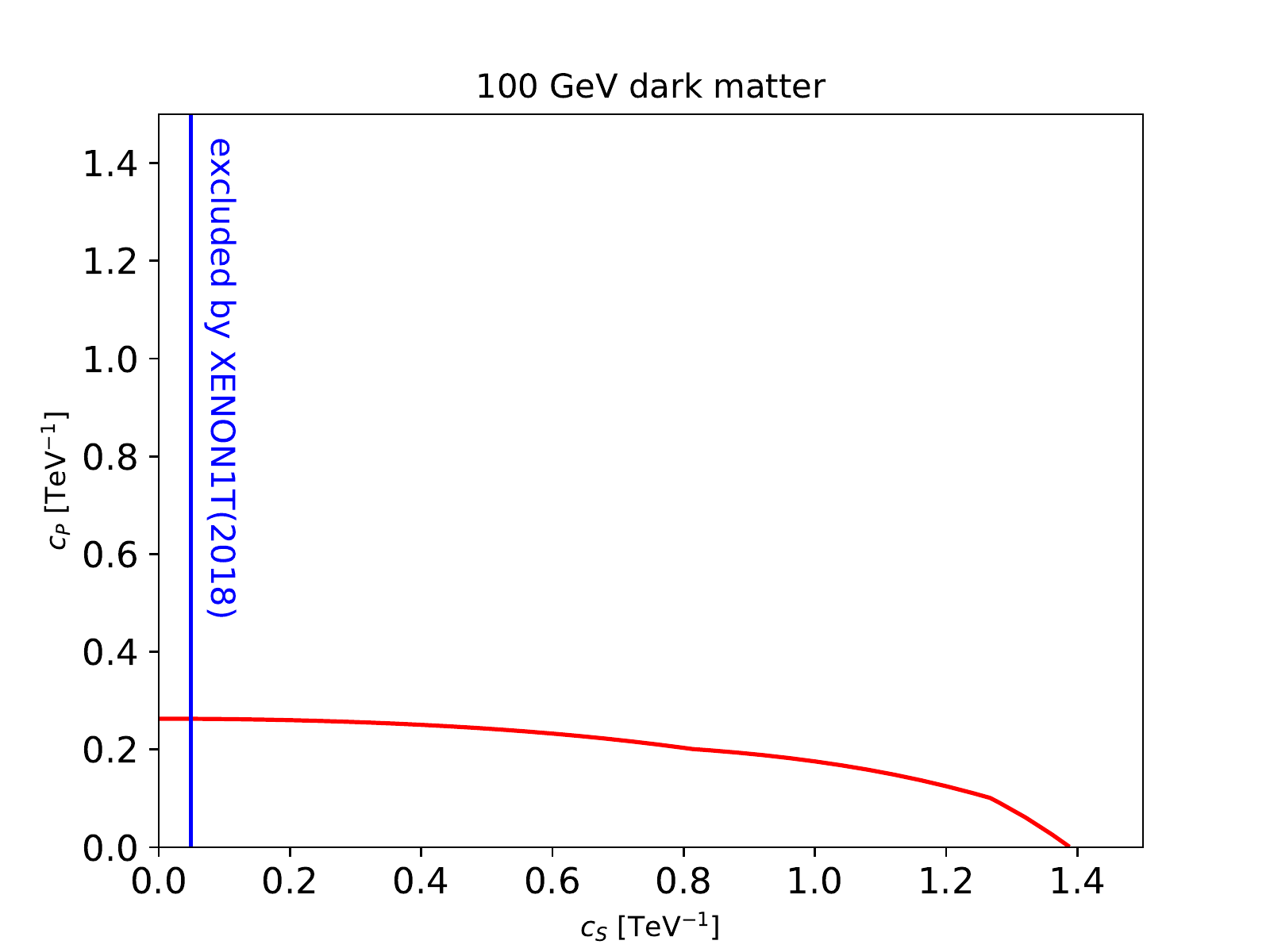}
\includegraphics[width=0.48\hsize]{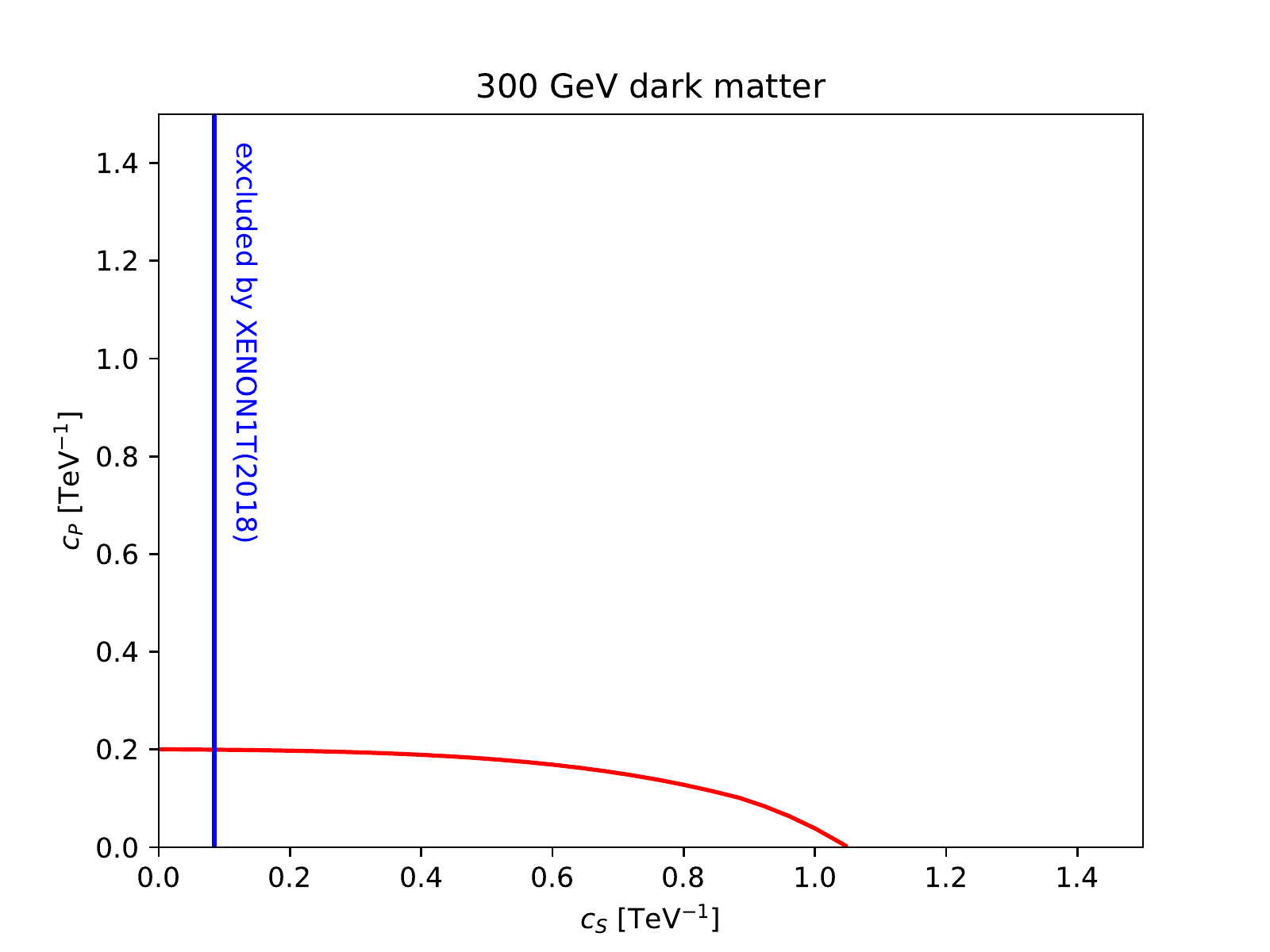}
\includegraphics[width=0.48\hsize]{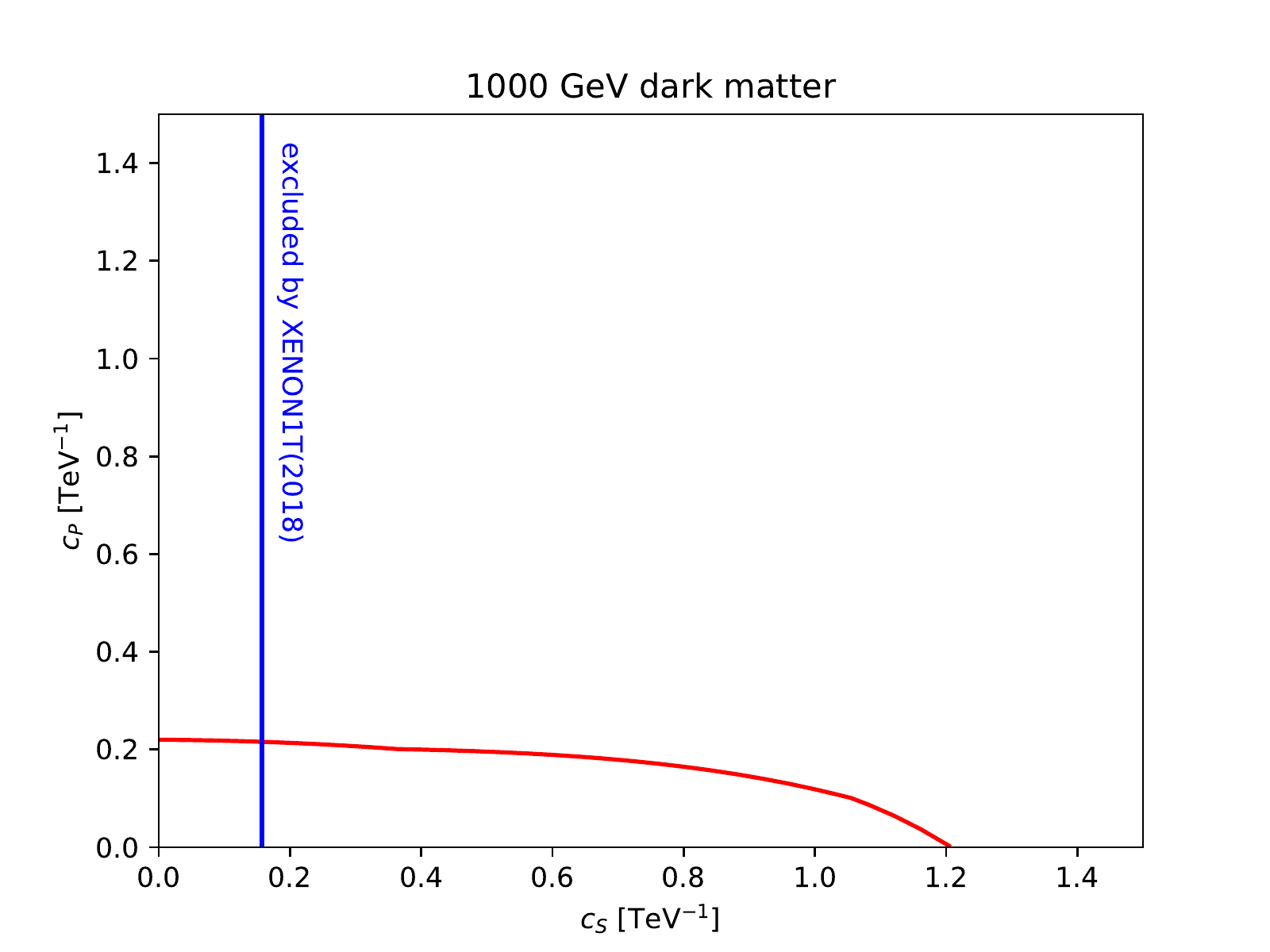}
\caption{
The fermion dark matter with $m_\text{DM} = 55$ (upper left), 100 (upper right), 300 (lower left), and 1000~GeV (lower right).
The relic abundance $\Omega h^2 = 0.1198$ 
on the red lines. The right region of the blue lines is excluded by the XENON1T experiment~\cite{1805.12562}.
}\label{fig:eff_CPodd}
\end{figure*}

Figure \ref{fig:eff_CPodd} shows the parameters which explain
the DM relic density $\Omega h^2 = 0.1198 \pm 0.0015$~\cite{1502.01589} in the $c_S$-$c_P$ plane.
We find the result is almost independent from $c_S$.
This is because the $CP$-conserving and the $CP$-violating operators contribute to
the DM annihilation processes by $p$ wave and $s$ wave, respectively, 
and thus the $CP$-violating operator plays the dominant role in the determination of 
the DM thermal relic abundance. 
We also find $c_P \sim 0.2$~TeV$^{-1}$ except for the Higgs funnel region.
This $c_P$ value implies 
that UV completions of the effective Lagrangian contain one or more new particles 
around ${\cal O}$(1)~TeV.

\begin{figure}[tb]
\includegraphics[width=0.6\hsize]{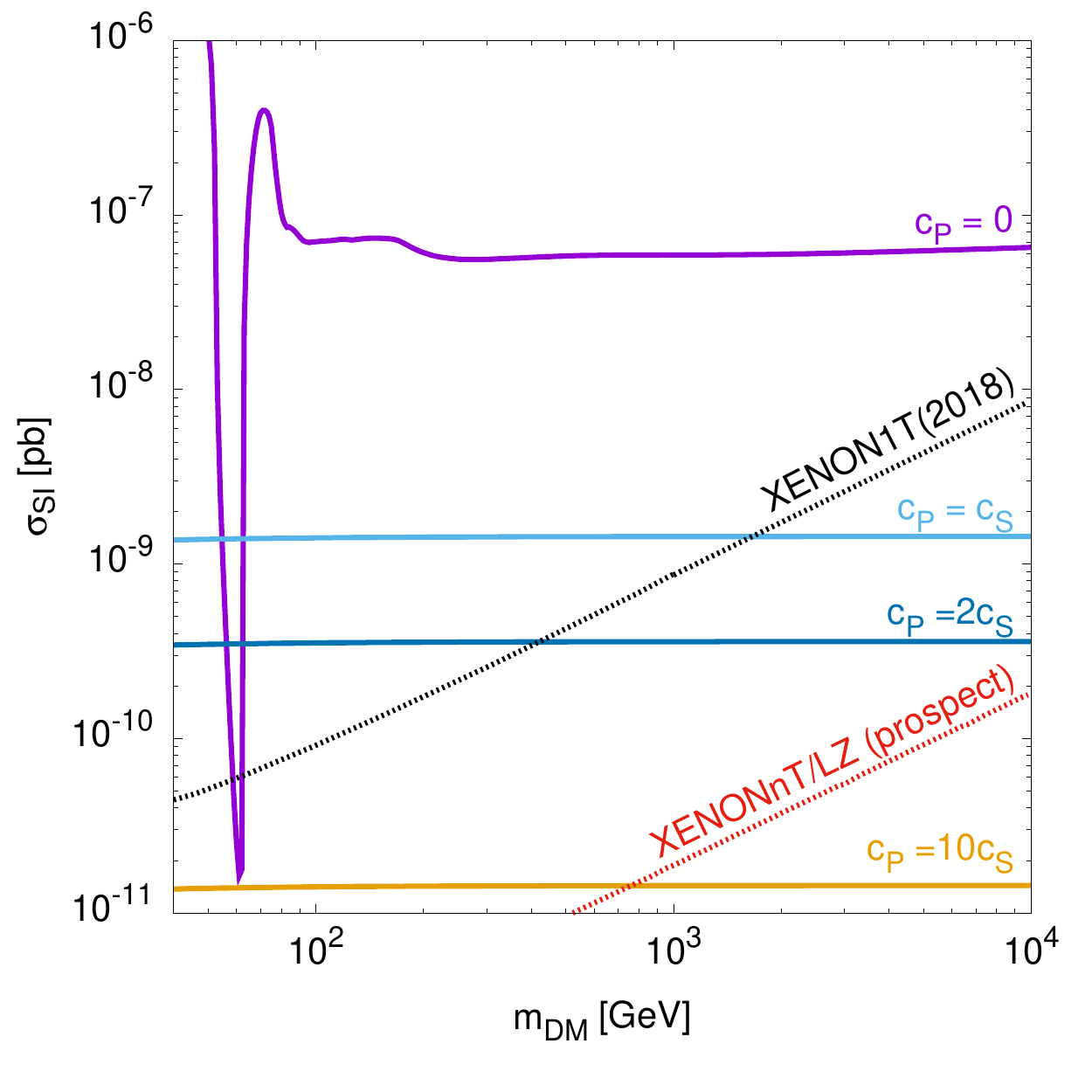} 
\caption{The spin-independent cross section for the DM-nucleon scattering.
The current upper bound on $\sigma_{\text{SI}}$ from the XENON1T experiment~\cite{1805.12562}
is shown by the black solid line. For $m_{\text{SM}}>$1~TeV, we extrapolate the result given in Ref.~\cite{1805.12562},
and show it by the black dashed line.
We also show the future prospects for XENONnT experiment~\cite{1512.07501} and LZ experiment~\cite{1802.06039}
by the red solid line. 
The other lines are the prediction of the effective theory given in Eq.~\eqref{eq:eff_lag}.
}
\label{fig:eff_only_CPeven}
\end{figure}

Figure \ref{fig:eff_only_CPeven} shows the spin-independent cross section for the
weakly interacting massive particle (WIMP)-nucleon scattering process as a function of the DM mass. Here we fix a parameter
so as to obtain the measured value of the DM energy density.
If the DM candidate interacts with the SM particles only through the $CP$-conserving operator,
namely $c_P = 0$, then 
the constraint from the DM direct detection experiments excludes the 
large part of the parameter space except for the Higgs funnel region.
On the other hand, if we have the $CP$-violating interaction ($c_P \neq 0$),
then it is easy to avoid the constraints from the current DM direct detection experiments 
as can be seen from the figure.
Therefore, we need nonzero $c_P$ except for the Higgs funnel region.

\begin{figure}[t]
\centering
\includegraphics[width=0.6\hsize]{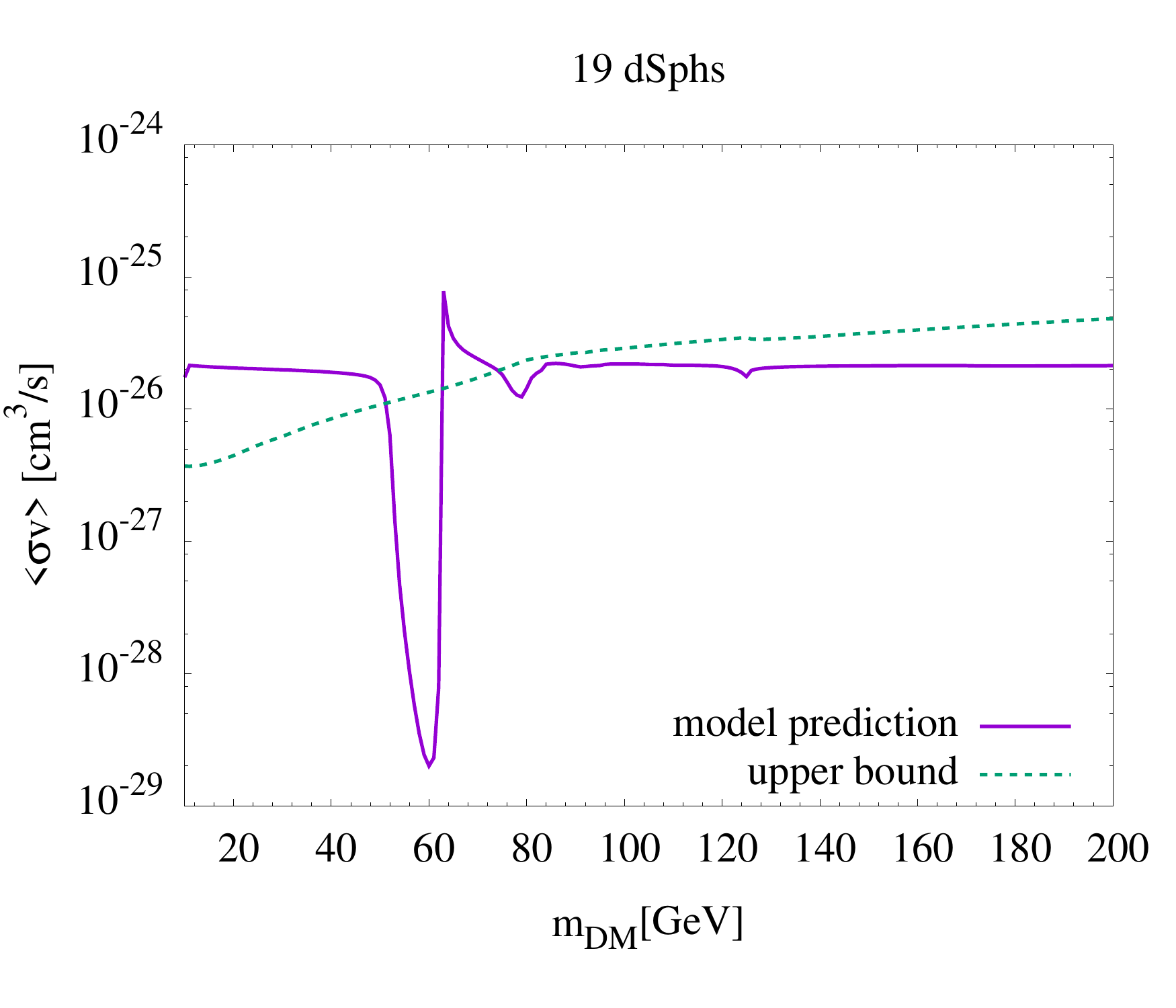} 
\caption{
The constraint on the effective theory from the measurements of the $\gamma$-ray flux from the
dSphs by the Fermi-LAT experiment.
In the upper panel, we use the 19 dSphs whose J-factors were measured kinematically.
}\label{fig:indirect_in_eff}
\end{figure}

The DM annihilation processes with $c_P$ are $s$ wave except for the Higgs funnel region, 
and thus the annihilation cross section at low temperature is not suppressed by powers of velocity.
This leads $\langle \sigma v \rangle \sim {\cal O}(10^{-26})$ cm$^3/s$ in the current 
Universe. As a result, we can expect the DM annihilation signals in the current Universe. 
We focus on $\gamma$-ray emissions by the DM annihilation from dSphs, 
which are regarded as good targets to observe such $\gamma$-ray emissions because of
the less astrophysical uncertainty.
The $\gamma$-ray flux from DM annihilation is given by~\cite{Fermi-LAT:2016uux}
\begin{align}
 \phi(\Delta \Omega, E_\text{min}, E_\text{max})
=
\frac{1}{4\pi} \frac{\langle \sigma v \rangle}{2 m_{\text{DM}}^2}
\int_{E_\text{min}}^{E_{\text{max}}}
\frac{d N_\gamma}{d E_\gamma}  dE_\gamma
\times J,
\label{eq:gamma-ray-flux}
\end{align}
where $J$ is the J-factor given by
\begin{align}
 J = \int_{E_\text{min}} \int_{\Delta \omega} \int_{l.o.s} \rho_{\text{DM}}^2 (\vec{r}(\ell)) d\ell d\Omega.
\end{align}
Each dSph has different J-factor, and some of their J-factors are already measured.
The variables other than the J-factor in Eq.~\eqref{eq:gamma-ray-flux} are calculable in particle physics.
We investigate the Fermi-LAT gamma-ray data~\cite{Fermi-LAT:2016uux},
and use the likelihood functions.\footnote{
The likelihood functions for each dSph are given at \url{http://www-glast.stanford.edu/pub_data/1203/}
.} 
We use the 19 dSphs whose J-factors were measured and are listed in the sixth column in Table 1 in Ref.~\cite{Fermi-LAT:2016uux}. 
With these J-factors, 
we use \texttt{micrOMEGAs}~\cite{1407.6129} to calculate the $\gamma$-ray flux from the DM annihilation.
Figure~\ref{fig:indirect_in_eff} shows the model prediction on $\langle \sigma v \rangle$ for $c_S = 0$ with the constraints from the indirect detection.
We find the upper bound on $\langle \sigma v \rangle$ and 
the regions for $m_{\text{DM}} <$ 50~GeV and for 62~GeV $< m_{\text{DM}} <$ 78~GeV are excluded.

There are two important points from 
Figs.~\ref{fig:eff_CPodd}--\ref{fig:indirect_in_eff}.
First, we need the $CP$-violating operator except for the funnel region. 
Second, the cutoff scale of this effective theory is around ${\cal O}(1)$~TeV.
This means a UV completion of this model requires one or more new particles at the TeV scale in addition to the DM particle.
In the next section, we focus on the singlet-doublet DM model~\cite{hep-ph/0510064, 0705.4493, 0706.0918}
 as a UV completion of the effective theory discussed in this section.

\section{The singlet-doublet model}\label{sec:f12}
\label{sec:F12}
In this section, we discuss the singlet-doublet model \cite{hep-ph/0510064, 0705.4493, 0706.0918}
as a UV completion of the effective theory which we have discussed in the previous section.
After reviewing the setup of the singlet-doublet model briefly, 
we discuss the constraint on the DM annihilation cross section from the Fermi-LAT experiment~\cite{Fermi-LAT:2016uux}.
Since the DM annihilation process is $s$ wave due to the $CP$-violating operator, 
the constraint is expected to be stronger than the case without $CP$ violation in the dark sector~\cite{1602.04788}.
The $CP$-violating operator also generates EDMs. 
We focus on the electron EDM and discuss its complementary
role to the DM direct detection searches.
We also investigate the stability of the Higgs potential. 
In the singlet-doublet model, the Higgs boson couples to the new fermion fields in the dark sector.
The couplings give the negative contribution to the beta function of the Higgs quartic coupling.
As a result, the Higgs potential becomes unstable compared to the Higgs potential in the SM.

Before starting to discuss the singlet-doublet model, 
we make brief comments on other UV completions. 
A model discussed in Refs.~\cite{1408.4929, 1701.04131, 1812.04092}, 
a gauge singlet fermion and a gauge singlet $CP$-odd scalar are introduced.
The singlet fermion is a DM candidate, and the $CP$-odd scalar is a mediator particle.
In the heavy $CP$-odd scalar mass limit, 
the model can be described by the effective theory discussed in Sec.~\ref{sec:eft}.
Another UV completion where the Higgs sector is also extended into the two-Higgs doublet model
was proposed in Ref.~\cite{1404.3716} and has been widely studied~\cite{1509.01110, 1611.04593, 1701.07427, 1705.09670, 1711.02110, 1712.03874, 1803.01574, 1804.02120, 1810.01039, 1810.09420}.
For other models, see Ref.~\cite{1506.04149}.

\subsection{The model}

We introduce
a gauge singlet Majorana fermion ($\omega$)
and an SU(2)$_L$ doublet Dirac fermion with hypercharge $Y = 1/2$
that consists of
a left-handed Weyl fermion ($\eta = (\eta^{+}, \eta^0)^T$) 
and a right-handed Weyl fermion ($\xi^\dagger = ( (\xi^{-})^\dagger, \xi^{0 \dagger})^T$);
see Table~\ref{tab:F12}. 
We impose a $Z_2$ symmetry on the model. 
Under the $Z_2$ symmetry, all the SM particles are even, and all the fermion fields we introduced in the above are odd. 
\begin{table}[htbp]
\centering
\caption{New particles in the Weyl notation.}
\label{tab:F12}
\begin{tabular}{|c|c|c|c|c|}\hline
              & SU(3)$_c$ & SU(2)$_L$ & U(1)$_Y$ & $Z_2$ \\ \hline \hline
$\omega$      &  1        &  1        &  0       &  $-1$ \\ \hline 
$\eta$        &  1        &  2        &  1/2     &  $-1$ \\ \hline
$\xi^\dagger$ &  1        &  2        &  1/2     &  $-1$ \\ \hline
\end{tabular}
\end{table}

The mass and Yukawa interaction terms for the $Z_2$-odd particles are given by
\begin{align}
 {\cal L}_{int.} 
=&
- \frac{M_1}{2} \omega \omega
- M_2 \xi \eta
- y \omega H^{\dagger} \eta
- y' \xi H \omega
+ (h.c.)
,
\end{align}
where $M_1$, $M_2$, $y$, and $y'$ are complex parameters. 
Three of the complex phases are unphysical because they can be absorbed by rotating the $Z_2$-odd fermion fields.
The physical phase ($CP$ phase) is given by
\begin{align}
 \phi = \text{Arg}(M_1^* M_2^* y y').
\end{align}
We introduce $r$ for later convenience,
\begin{align}
 r = \frac{|y|}{|y'|}.
\end{align}
We have five free parameters, $(|M_1|, |M_2|, \sqrt{|y|^2+|y'|^2}, r, \phi)$.
We determine $\sqrt{|y|^2+|y'|^2}$ so as to realize the measured value of the dark matter energy density, 
$\Omega h^2 = 0.1198 \pm 0.0015$~\cite{1502.01589}.
We use \texttt{micrOMEGAs}~\cite{arXiv:1407.6129} to calculate the dark matter relic abundance. 
After the Higgs field develops the vacuum expectation value, the singlet field and the doublet field 
are mixed.
As a result, there are three Majorana fermions ($\chi^0_{1,2,3}$) 
and a pair of charged fermions ($\chi^\pm$) 
in the $Z_2$-odd sector. The lightest Majorana fermion is the DM candidate in this setup.

\subsection{Relation to the effective theory}
In order to see the relation of this model to the effective theory, we
discuss the large $M_2$ regime.
If $M_2 \gg v, M_1$, 
then we can integrate out the doublet fields, and the model is reduced into the effective theory.
Up to dimension-6 operators, we find the following terms in the effective Lagrangian.
\begin{align}
 {\cal L} 
=&
 {\cal L}_{\text{SM}}
+
\frac{1}{2} 
 \Psi_s  \left(  i \gamma^{\mu} \partial_{\mu} - |M_1|  \right) \Psi_s
\nonumber\\
&
+
\text{Re} \left( \frac{|yy'| e^{i\phi}}{|M_2|}\right)
H^{\dagger} H \bar{\Psi}_s \Psi_s
-
\text{Im} \left( \frac{|yy'| e^{i\phi}}{|M_2|}\right)
H^{\dagger} H \bar{\Psi}_s i \gamma^5 \Psi_s
\nonumber\\
&
-
\frac{|y|^2 - |y'|^2}{4 |M_2|^2}
\left(
(H^{\dagger} i \tensor{D}_{\mu} H)
\bar{\Psi}_s \gamma^{5}\gamma^{\mu} \Psi_s
\right)
+
\frac{|y|^2 + |y'|^2}{2 |M_2|^2}
\left(
H^{\dagger} H
\bar{\Psi}_s \gamma^{\mu} i \partial_{\mu} \Psi_s
\right)
\\
=&
{\cal L}_{SM}
+
\frac{1}{2} 
 \Psi_s  \left(  i \gamma^{\mu} \partial_{\mu} - |M_1|  \right) \Psi_s
\nonumber\\
&
+
\left(
\text{Re} \left( \frac{|yy'| e^{i\phi}}{|M_2|}\right)
+
M_1 \frac{|y|^2 + |y'|^2}{2 M_2^2}
\right)
H^{\dagger} H \bar{\Psi}_s \Psi_s
-
\text{Im} \left( \frac{|yy'| e^{i\phi}}{|M_2|}\right)
H^{\dagger} H \bar{\Psi}_s i \gamma^5 \Psi_s
\nonumber\\
&
-
\frac{|y|^2 - |y'|^2}{4 |M_2|^2}
(H^{\dagger} i \tensor{D}_{\mu} H)
\bar{\Psi}_s \gamma^{5}\gamma^{\mu} \Psi_s
,
\label{eq:F12_eff}
\end{align}
where
\begin{align}
 \Psi_s
=&
\left(
\begin{matrix}
 \omega \\ \omega^{\dagger}
\end{matrix}
\right)
,\\
A \tensor{\partial_{\mu}} B 
=& A \partial_{\mu} B - (\partial_{\mu} A) B
.
\end{align}
Here we have used the equation of motion for $\Psi_s$.
Comparing Eq.~\eqref{eq:F12_eff} with Eq.~\eqref{eq:eff_lag}, we find that $M_2$ plays the role of the cutoff scale in the effective theory.

As can be seen from Eq.~\eqref{eq:F12_eff},
there is a region in the parameter space 
where the scalar coupling is highly suppressed or even vanishes 
while keeping the pseudoscalar coupling.
This accidental cancellation of the scalar coupling is known as the blind spot~\cite{1311.5896, 1411.1335, 1505.03867, 1509.05323, 1603.07387}. 
Therefore the scalar coupling can be parametrically suppressed in this model.

The last term in Eq.~\eqref{eq:F12_eff} is absent in Eq.~\eqref{eq:eff_lag} because we did not consider the effects of dimension-6 operators in the previous section.
Since it generates the DM-$Z$ coupling, it affects the DM annihilation cross section and also contributes to the spin-dependent DM-nucleon scattering process.
If $|y| = |y'|$, namely $r=1$, this term vanishes
because of the symmetry enhancement in the dark sector. If $r=1$, the dark sector is symmetric under the exchange of $\eta$ with $\xi$, and consequently the DM-$Z$ coupling vanishes.
The beta function of $r$ at the one-loop level is given by
\begin{align}
 \mu \frac{d}{d\mu} \frac{y}{y'}
=&
 -\frac{1}{(4\pi)^2} \frac{3y}{2y'} \left( |y|^2  - |y'|^2\right)
.
\end{align}
It is clear that $r=1$ is a fixed point because the beta function is 0.
This is another way to see the symmetry enhancement at $r=1$. 
In the following numerical analysis, we take $r=1$ for simplicity.

\subsection{Electron EDM}

EDMs, $d_f$, are sensitive to $CP$ violation
and are defined through
\begin{align}
 {\cal H}_{eff}
=&
 i \frac{d_f}{2}
\bar{\psi}_{f}
\sigma_{\mu \nu} \gamma_5
\psi_{f}
F^{\mu \nu}
.
\end{align}
As we have discussed in Sec.~\ref{sec:eft}, $CP$ violation is important to avoid the constraints from
the DM direct detection experiments, and thus the EDMs are naturally expected in this model~\cite{hep-ph/0510064, 0705.4493, 1411.1335, 1702.07236}.
In particular, we focus on the eEDM, $d_e$, that is severely constrained by experiments.

The current bound on the eEDM is given by the ACME experiment~\cite{Andreev:2018ayy},
$|d_e| \leq 1.1 \times 10^{-29}~e$ cm (90\% C.L.).
There are some prospects for eEDM~\cite{1208.4507,Kawall:2011zz}, 
and we can expect that the eEDM is detectable in future if $|d_e| \gtrsim {\cal O}(10^{-30})~e$ cm. 
We use $|d_e| = 10^{-30}~e$ cm as the prospect of eEDM in the following analysis.

The eEDM in this model is given by~\cite{1411.1335, 1702.07236} 
\begin{align}
 \frac{d_e}{e}
=&
- 
\frac{2 \alpha}{(4\pi)^3 s_W^2} \sqrt{2} G_F
m_{\chi^{\pm}} m_e
\sum_{j=1}^3
\text{Im}(V_{2j} V_{3j})
m_{\chi^0_j}
{\cal I}_j
,
\end{align}
where $m_{\chi^\pm}$ is the mass of the charged $Z_2$-odd fermion, which is the same
as $|M_2|$ at the tree level,
$V_{ij}$ is a 3 by 3 matrix to connect the gauge eigenstates and the 
mass eigenstates of the $Z_2$-odd fermions,
\begin{align}
 \begin{pmatrix} \omega \\ \eta^0 \\ \xi^0  \end{pmatrix}
=
 \begin{pmatrix}
  V_{11} & V_{12} & V_{13} \\
  V_{21} & V_{22} & V_{23} \\
  V_{31} & V_{32} & V_{33}
 \end{pmatrix}
 \begin{pmatrix} \chi^0_1 \\ \chi^0_2 \\ \chi^0_3  \end{pmatrix}
,
\end{align}
and
\begin{align}
{\cal I}_j
=&
 \int_0^1 dz
\frac{1-z}{m_{\chi^{\pm}}^2 (1-z) + m_{\chi^0_j}^2 z - m_W^2 z (1-z)}
\ln \frac{ m_{\chi^{\pm}}^2 (1-z) + m_{\chi^0_j}^2 z}{m_W^2 z (1-z)}
.
\end{align}
For $|M_2| \gg |M_1|, v$, 
\begin{align}
 \frac{d_e}{e}
\simeq&
- \frac{1}{2}
\frac{1}{(4\pi)^4}
\left(\frac{e}{s_W} \right)^2
 m_e
 \frac{|M_1 M_2 y y'| \sin\phi}{|M_2|^4}
 \ln \frac{|M_2|^2}{m_W^2}
.
\label{eq:EDM}
\end{align}
It is clear that the eEDM is proportional to Im$(M_1^* M_2^* y y')$.
For $|M_2| \gg |M_1|, v$, we find $m_{\text{DM}} \simeq M_1$, 
and thus the EDMs are proportional to the dark matter mass.

\subsection{Stability of the Higgs potential}
Before studying the DM physics, we discuss the stability of the Higgs potential.\footnote{The stability without $CP$ violation in the singlet-doublet model is discussed in Refs.~\cite{1203.5106, 1708.01614, 1811.08743}.
}
It is worse than the SM case
because of the Higgs couplings to the fermion fields in the dark sector~\cite{1203.5106}.
The beta function of the Higgs quartic coupling at the one-loop level is given by
\begin{align} 
(4\pi)^2 \mu \frac{d\lambda}{d \mu}
=&  
+24 \lambda^{2}
+\frac{3}{8} g_{1}^{4} +\frac{3}{4} g_{1}^{2} g_{2}^{2} +\frac{9}{8} g_{2}^{4} -3 g_{1}^{2} \lambda -9 g_{2}^{2} \lambda 
 \nonumber\\
&+4 \lambda (|y|^2+|y'|^2) -2 (|y|^{2} + |y'|^{2})^2  \nonumber \\ 
&+12 \lambda y_t^2 -6 y_t^4,
\label{eq:rge_lambda}
\end{align}
where $\lambda$ is defined via the Higgs potential,
\begin{align}
 V =& m^2 H^\dagger H + \lambda (H^\dagger H)^2.
\end{align}
As can be seen from Eq.~\eqref{eq:rge_lambda},
the stability gives severe constraint for large $y$ and $y'$ regime.

To estimate the stability bound of the Higgs potential,
we define the cutoff scale $\Lambda$ by
\begin{align}
 \lambda(\Lambda) = 0.
\end{align}
The Lagrangian of the singlet-doublet model is no longer valid above the scale of $\Lambda$.
Otherwise the lifetime of our vacuum would be shorter than the age of the Universe.
We derive the beta function of $\lambda$ up to the two-loop level by
using \texttt{SARAH}~\cite{Staub:2008uz}.
We use the SM beta functions below $M_2$,
and we change the beta function at $\mu = M_2$.
We numerically solve the renormalization group equations at the two-loop level.
We calculate the threshold correction at the one-loop level.
We find
\begin{align}
 \lambda(M_2+0) - \lambda(M_2-0)
\simeq&
- \frac{1}{3 v^2}
\left(
\Sigma_{\text{1PI}} + 2 \lambda v {\cal T}
\right)
+
\frac{\lambda}{v^2}
\frac{4}{g^2} \Sigma_{WW}^{\text{BSM}}(0)
,
\end{align}
where $\Sigma_{\text{1PI}}$, ${\cal T}$, and $\Sigma_{WW}^{\text{BSM}}$ are given in Appendix~\ref{app:threshold}.
The details of the derivation are also given in Appendix~\ref{app:threshold}.
The SM input parameters are given in~\cite{1307.3536},
\begin{align}
 g'(m_t) =& 0.35830, \quad
 g(m_t) = 0.64779, \quad
 g_s(m_t) = 1.1666, \\
 y_t(m_t) =& 0.93690, \quad
 \lambda(m_t) = 0.12604, \quad
 m_t = 173.34~\text{GeV}.
\end{align}

From the analysis of the effective theory, we know that 
the dimension-5 operator suppressed by $\Lambda = 5$~TeV affects the dark matter annihilation cross section. 
We require $\Lambda > 50$~TeV  (500~TeV) to make the uncertainty coming from the higher dimension operators
smaller than 10\% (1\%).

\subsection{Indirect detection}
Some region of the parameter space where $ 0 < \phi < \pi $ can be constrained by DM
indirect detection because the DM annihilation process is $s$ wave in the $CP$-violating region.
As discussed in Sec.~\ref{sec:eft}, the $CP$-violating interaction term is important 
to avoid the constraints from the DM direct detection experiments. 
Thus studying the constraints from indirect detection searches are complementary to
the constraints from the direct detection experiments.
In the following, we focus on the $\gamma$-ray emission by the DM annihilation as in Sec.~\ref{sec:eft}.

\subsection{Higgs invisible decay}
The Higgs boson can decay into DM pairs if it is heavier than twice the DM mass.
This decay mode can be observed as an invisible decay at the LHC.
The ATLAS and CMS experiments give the current upper bound on the Higgs invisible branching ratio as
Br($H \to \text{invisible}) < 0.26$ at 95\% C.L.~\cite{ATLAS-CONF-2018-054, 1809.05937}.
We use this bound for $m_{\text{DM}} < m_h/2 \simeq 63$~GeV.

\subsection{Constraints from direct, indirect, eEDM experiments, and stability bound}
We show the constraints from the direct and indirect detection experiments, the eEDM experiment, and stability bound.
Some prospects are also discussed.

\begin{figure}[tb]
\includegraphics[width=0.45\hsize]{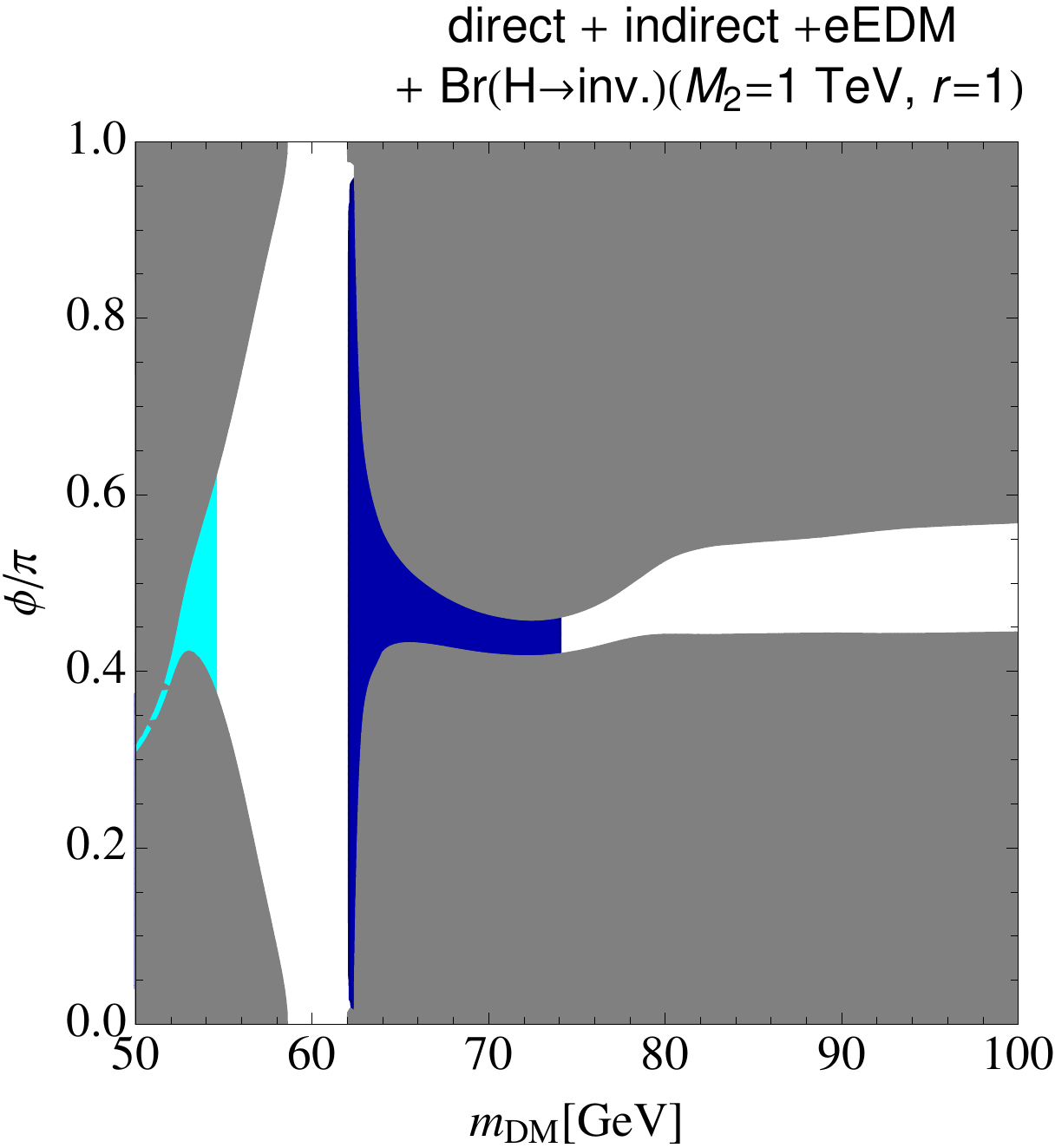}
\qquad
\includegraphics[width=0.45\hsize]{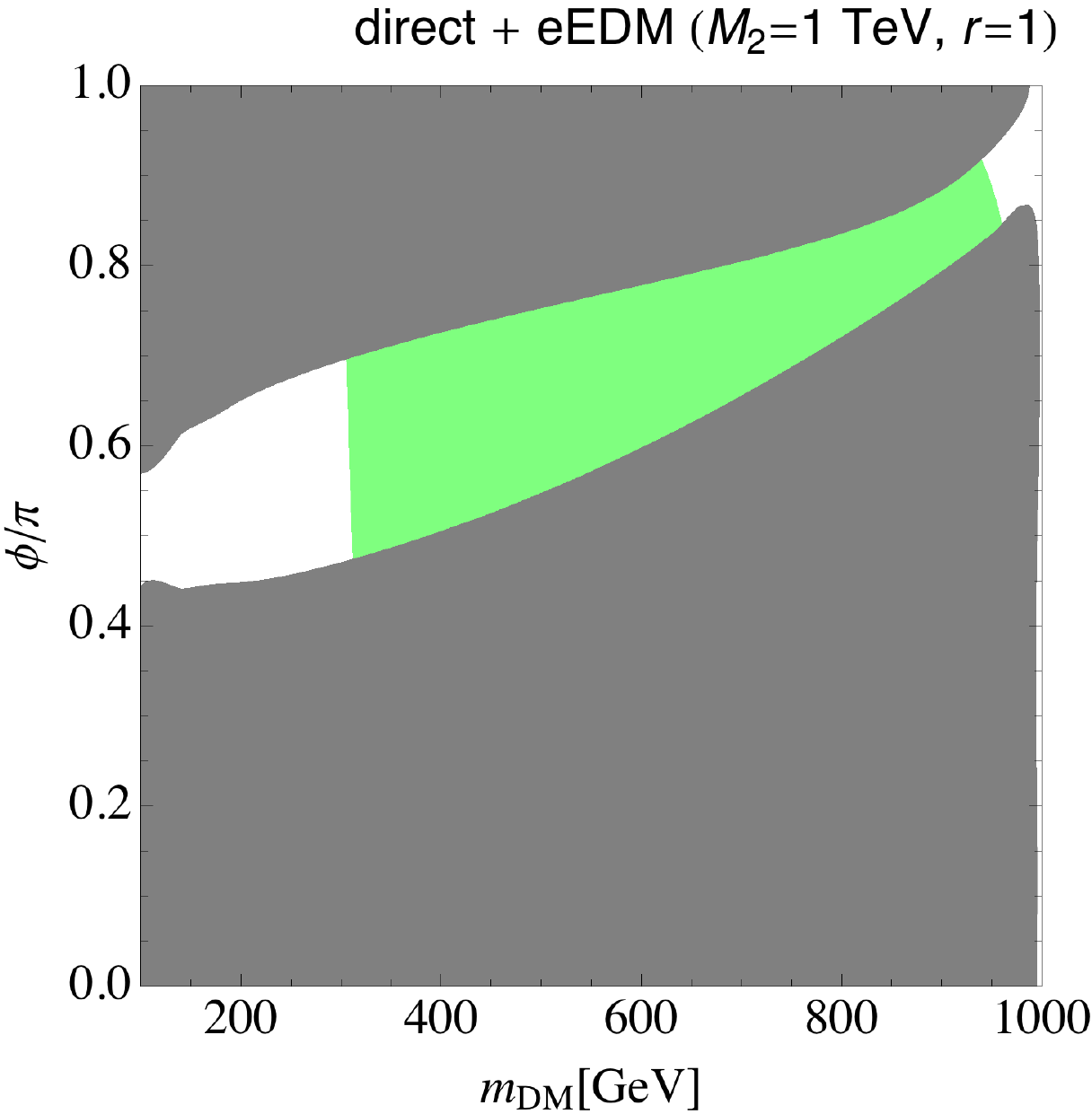}
\\
\includegraphics[width=0.45\hsize]{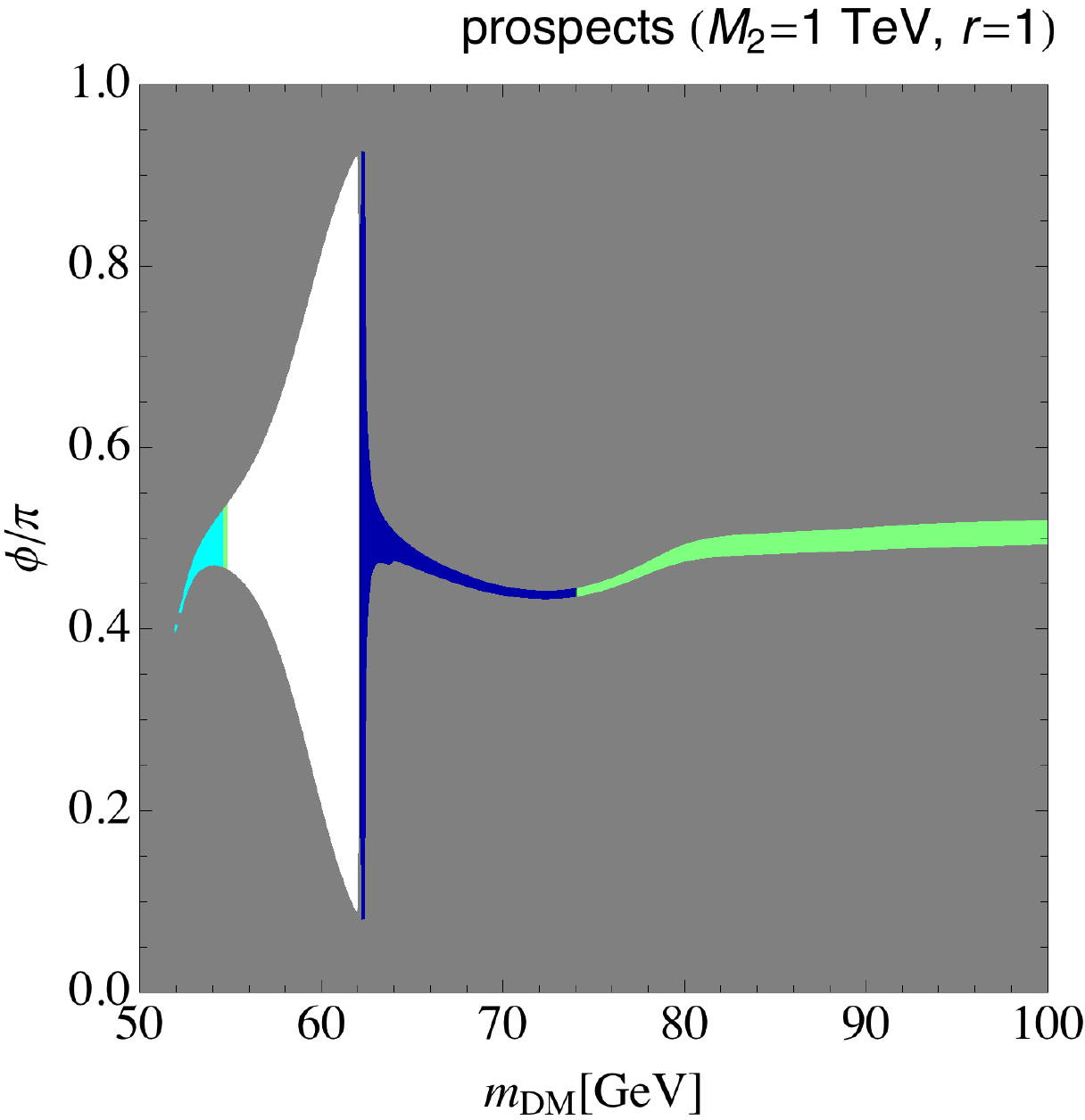}
\qquad
\includegraphics[width=0.45\hsize]{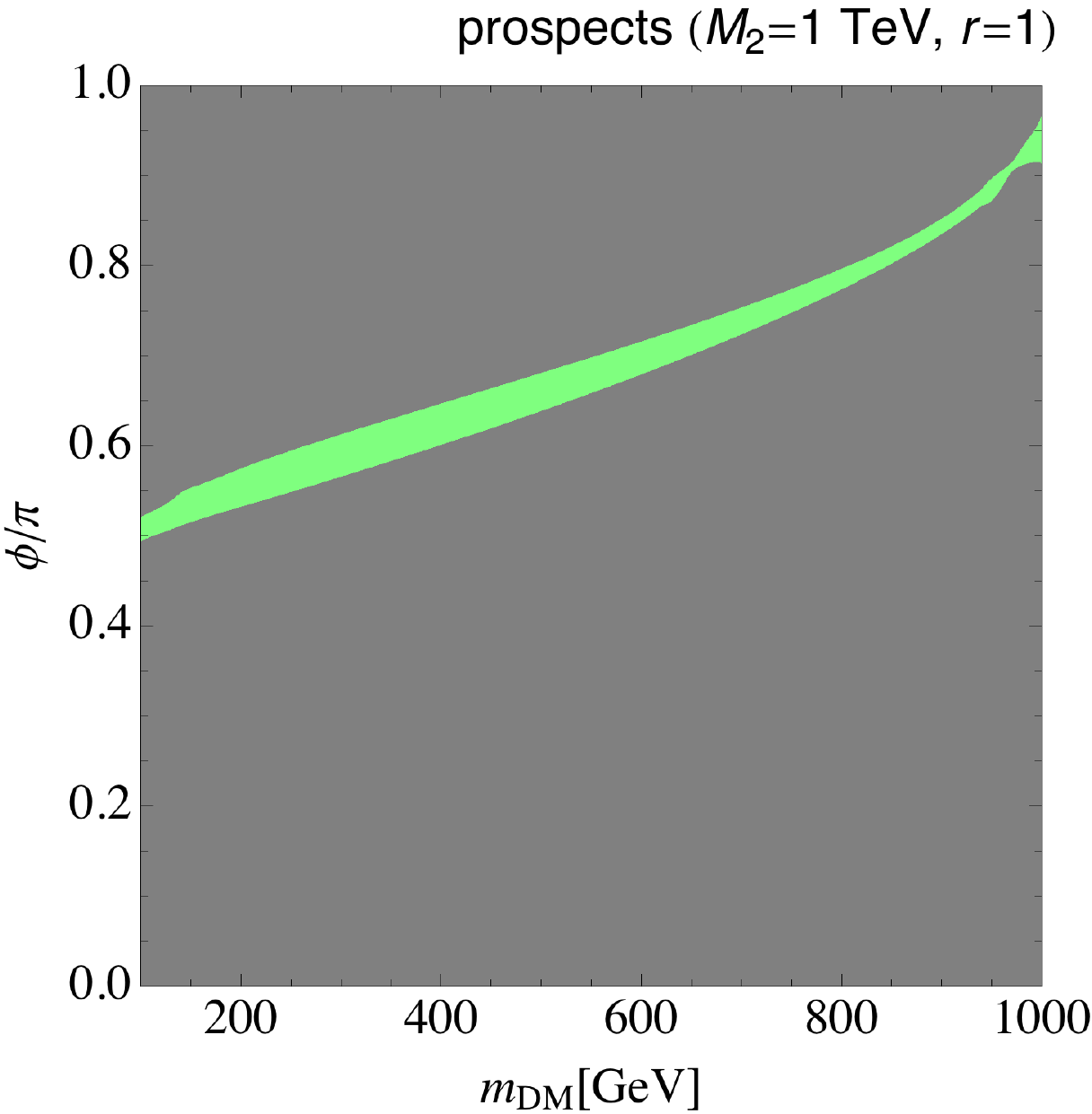}
\caption{ 
Current situation (upper panels) and prospects (lower panels)of the singlet-doublet model
for $M_2 = 1$~TeV and $r=1$.
The gray, green, and blue shaded regions are excluded by the XENON1T~\cite{1805.12562}, 
ACME~\cite{Andreev:2018ayy}, and Fermi-LAT experiment~\cite{Fermi-LAT:2016uux},
respectively.
The cyan region is excluded by the measurement of the Higgs invisible decay at the LHC experiments~\cite{ATLAS-CONF-2018-054, 1809.05937}.
For the prospect, we used LZ~\cite{1802.06039} and $|d_e| = 10^{-30}~e$ cm~\cite{1208.4507,Kawall:2011zz}.
}
\label{fig:result_M2=1TeV}
\end{figure}
The upper panels in Fig.~\ref{fig:result_M2=1TeV} show the current status
of the singlet-doublet model.
The constraint from the XENON1T experiment is very strong.
If the model respects $CP$ invariance in the dark sector, only a tiny region 
around $m_{\text{DM}} \sim m_h/2$, the so-called Higgs funnel region, 
is consistent with the XENON1T experiment.
On the other hand, if $CP$ is not a good symmetry in the dark sector, namely
$0 < \phi < \pi$, then the wide range of the dark matter mass is possible
 thanks to the the pseudoscalar interaction associated with
the $CP$ violation in the dark sector. Therefore, varying the $CP$ phase
is important in the analysis of the singlet-doublet model.
The constraint from the ACME experiment excludes 300~GeV~$\lesssim m_{\text{DM}}\lesssim 950$~GeV.
For $m_{\text{DM}} \gtrsim 950$~GeV, smaller Yukawa couplings can realize the thermal relic abundance thanks to coannihilation processes, and thus eEDM is smaller than the other region.
Since the eEDM is proportional to $m_{\text{DM}}$, as shown in Eq.~\eqref{eq:EDM},
the lighter mass is less constrained.
The constraint from the gamma-ray flux from the dSphs excludes $m_h/2 < m_{\text{DM}} < 74$~GeV
at 90\% confidence level.
The constraint from the Higgs invisible decay search excludes $m_{\text{DM}}<54$~GeV at 95\% confidence level.

The lower panels in Fig.~\ref{fig:result_M2=1TeV} show the prospect of the model.
Here we use the LZ experiment~\cite{1802.06039}
and $|d_e|< 10^{-30}~e$ cm~\cite{1208.4507,Kawall:2011zz} for the prospect
of the direct detection experiment and the eEDM measurements, respectively.
We find that the $CP$-conserving region can be excluded if there is no signal from
the direct detection experiment in future. In the $CP$-conserving case, the DM pair
annihilates into $b\bar{b}$ through the Higgs exchange in the s-channel 
in the Higgs funnel region, and it is $p$ wave. On the other hand, it is $s$ wave
in the $CP$-violating case. Therefore, larger Yukawa couplings are required for the $CP$-conserving case
to explain the measured value of the dark matter energy density,
and thus the $CP$-conserving regions receive the stronger constraint than the $CP$-violating region
from the direct detection experiment.
The eEDM can cover most of the $CP$-violating region except the Higgs funnel region.
In conclusion, if there is no signal from the direct detection experiments nor eEDM,
the singlet-doublet model is consistent only at the Higgs funnel region.
We address the Higgs funnel region later.

\begin{figure}[tb]
\includegraphics[width=0.45\hsize]{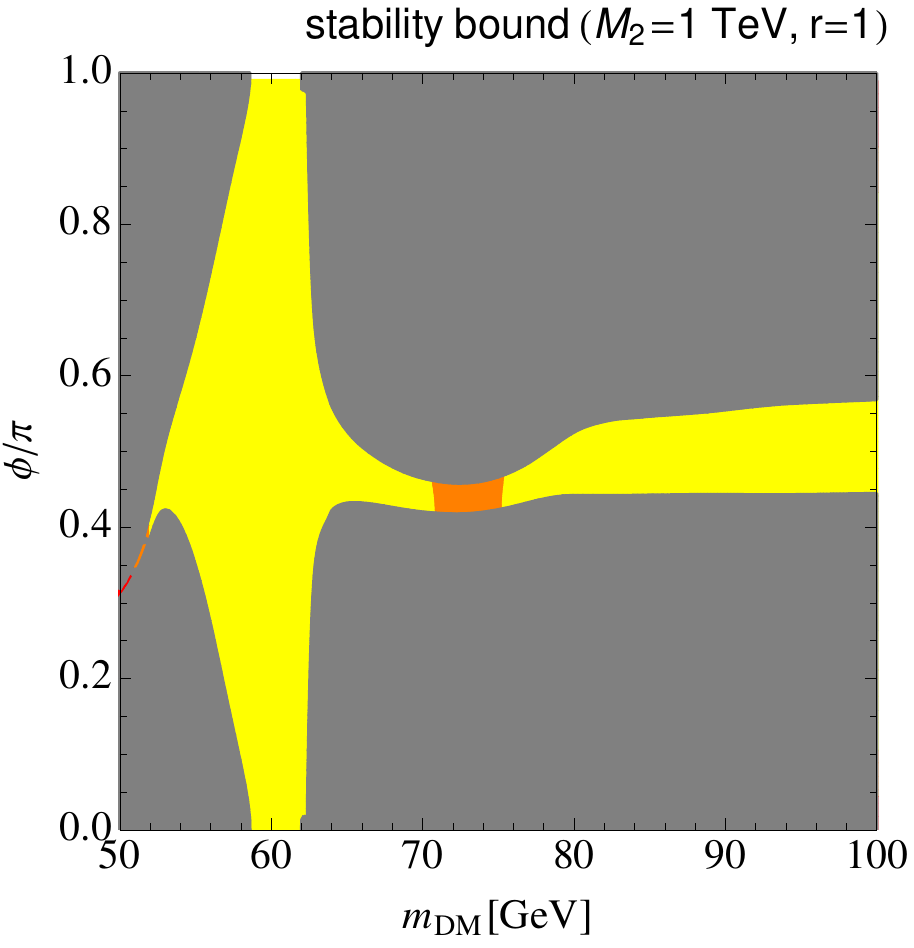}
\qquad
\includegraphics[width=0.45\hsize]{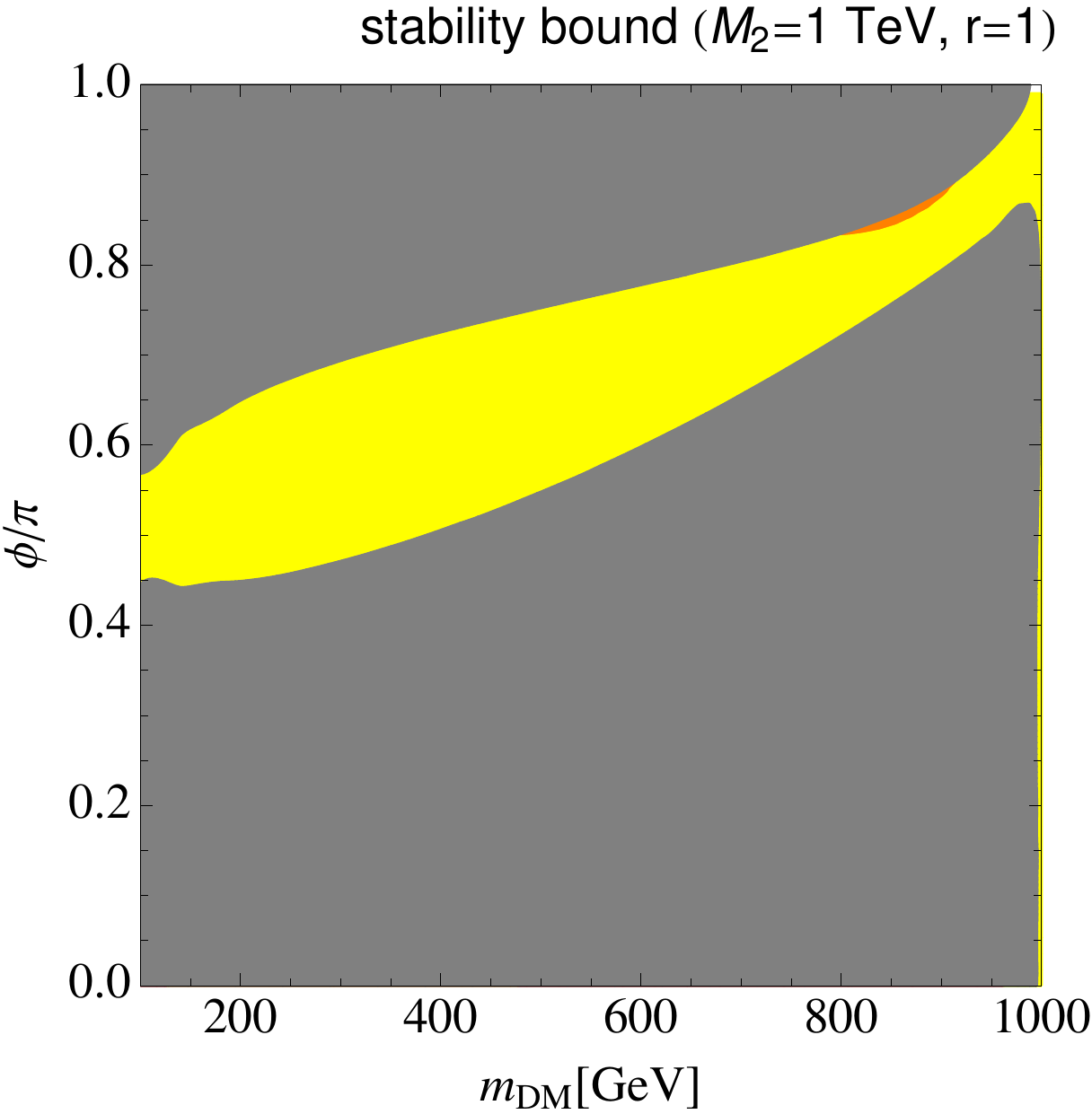}
\\
\includegraphics[width=0.45\hsize]{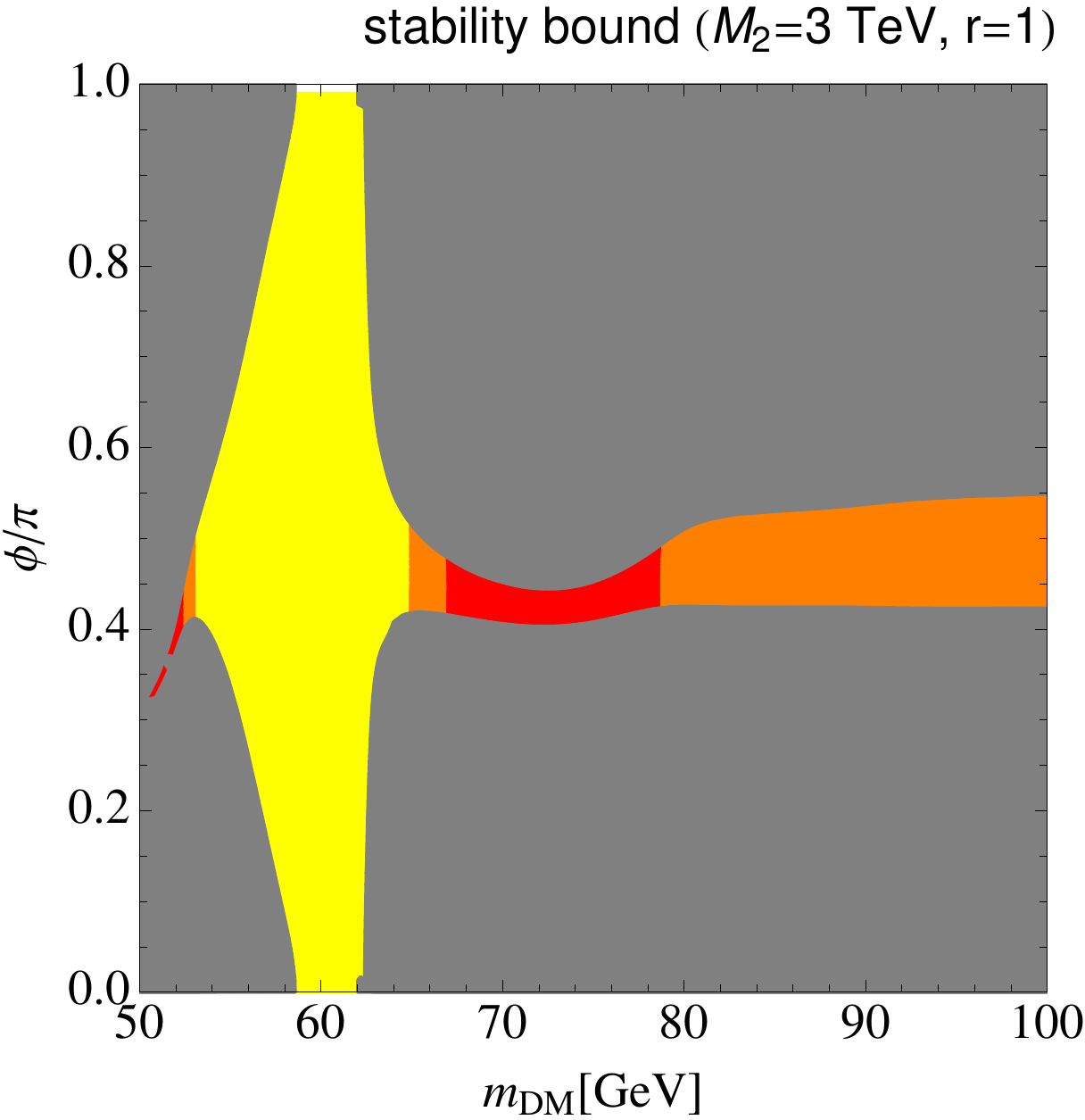}
\qquad
\includegraphics[width=0.45\hsize]{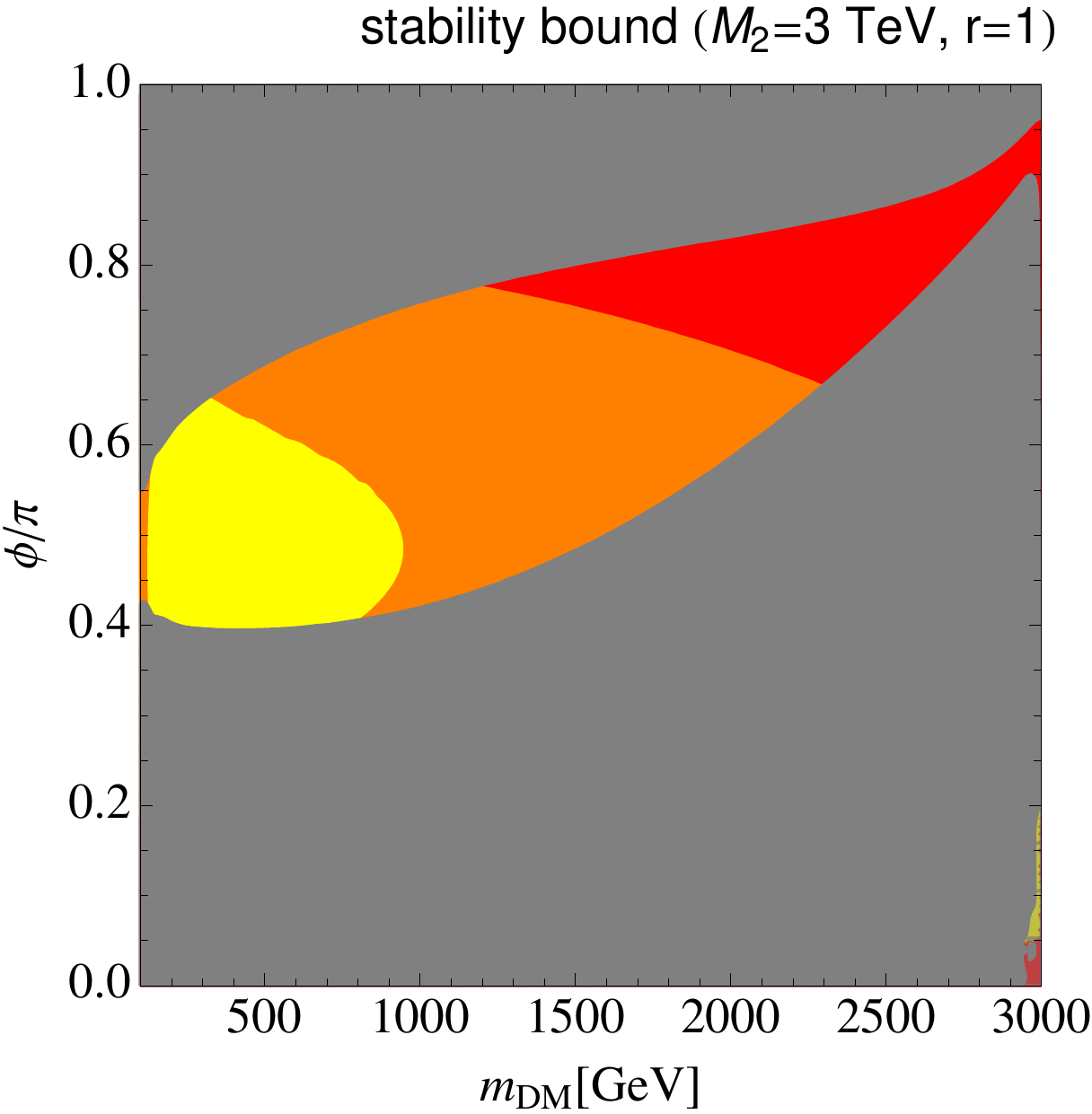}
\caption{ 
The cutoff scale from the stability bound.
The upper (lower) panels are for $M_2 = 1 (3)$~TeV and $r=1$.
In the red, orange, and yellow regions,
$\Lambda < 50$~TeV, 50~TeV$< \Lambda < 500$~TeV, and 500~TeV$< \Lambda$, respectively.
The gray regions are excluded by the XENON1T experiment~\cite{1805.12562}. 
}
\label{fig:result_cutoff}
\end{figure}
The upper panels in Fig.~\ref{fig:result_cutoff} show the stability bound.
We find $\Lambda > 500$~TeV in most of the region of the parameter space.
In general, UV physics above the scale $\Lambda$ gives higher dimensional operators to the Lagrangian of the singlet-doublet model.
There is a theoretical uncertainty in this sense,
and if $\Lambda = 5$~TeV, then a theoretical uncertainty from the higher dimensional operators in our calculations is not negligible
because $c_P$ in Sec.~\ref{sec:eff} is comparable to $\Lambda^{-1} = (5~\text{TeV})^{-1}$.
If $\Lambda > 500$~TeV, the uncertainty becomes less than 1\%. 
Since the energy density of the dark matter energy density is determined with 1\% accuracy~\cite{1502.01589},
we can neglect the uncertainty of ignoring the higher dimension operator from UV physics for $\Lambda > 500$~TeV.
Thus, our result shown in Fig.~\ref{fig:result_M2=1TeV} does not receive uncertainty from the higher dimension operators.

\begin{figure}[tb]
\includegraphics[width=0.45\hsize]{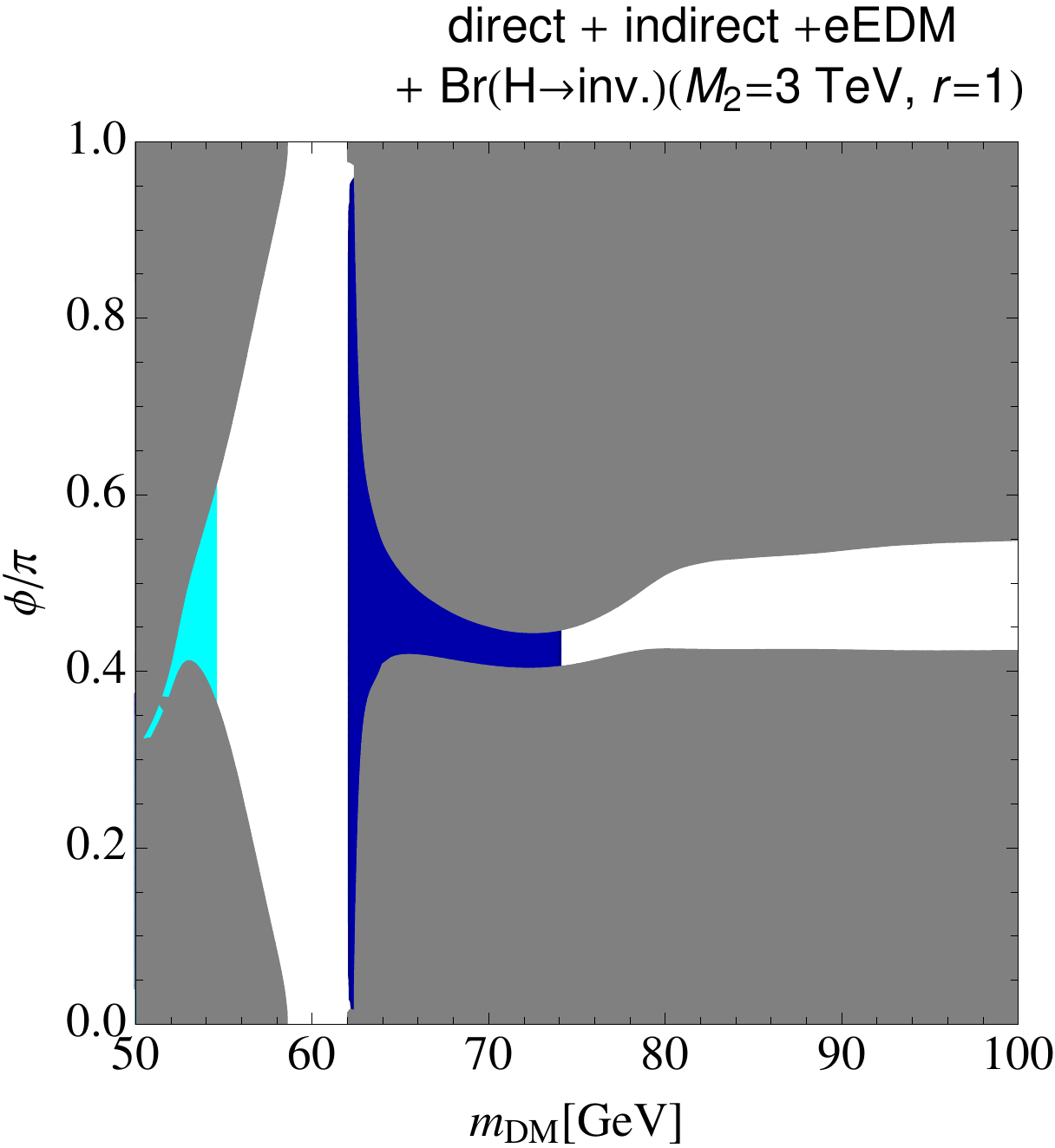}
\qquad
\includegraphics[width=0.45\hsize]{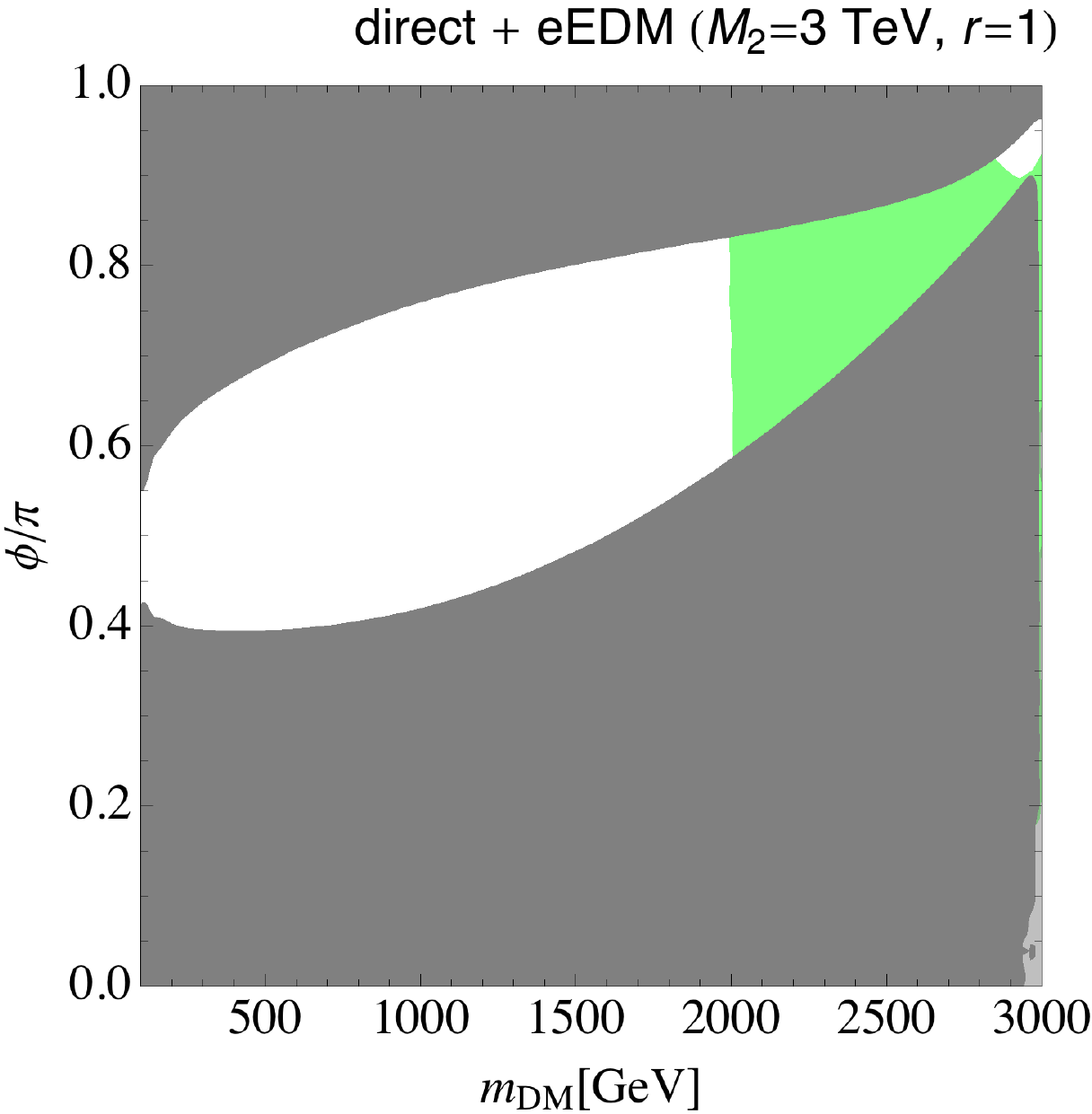}
\\
\includegraphics[width=0.45\hsize]{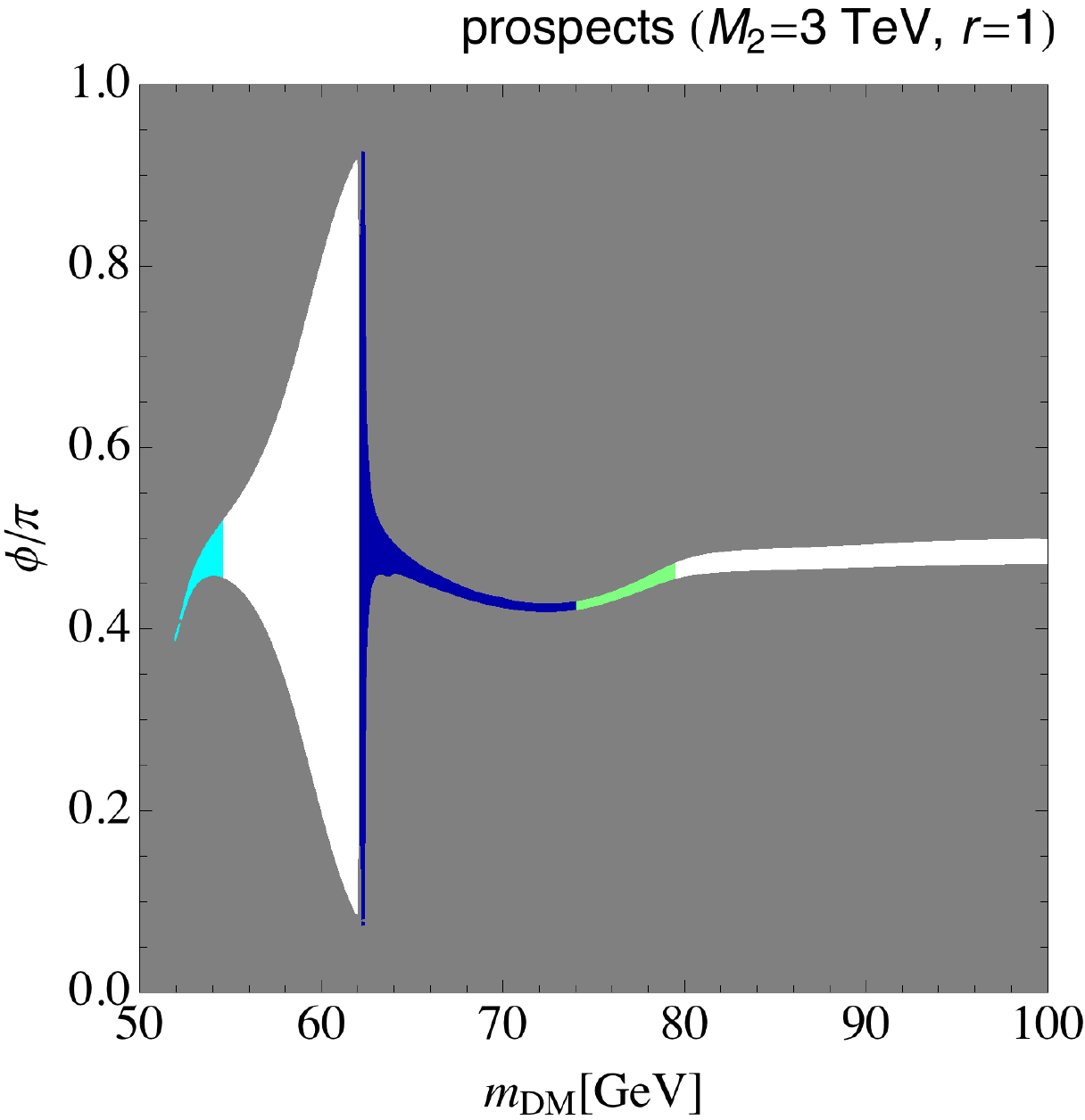}
\qquad
\includegraphics[width=0.45\hsize]{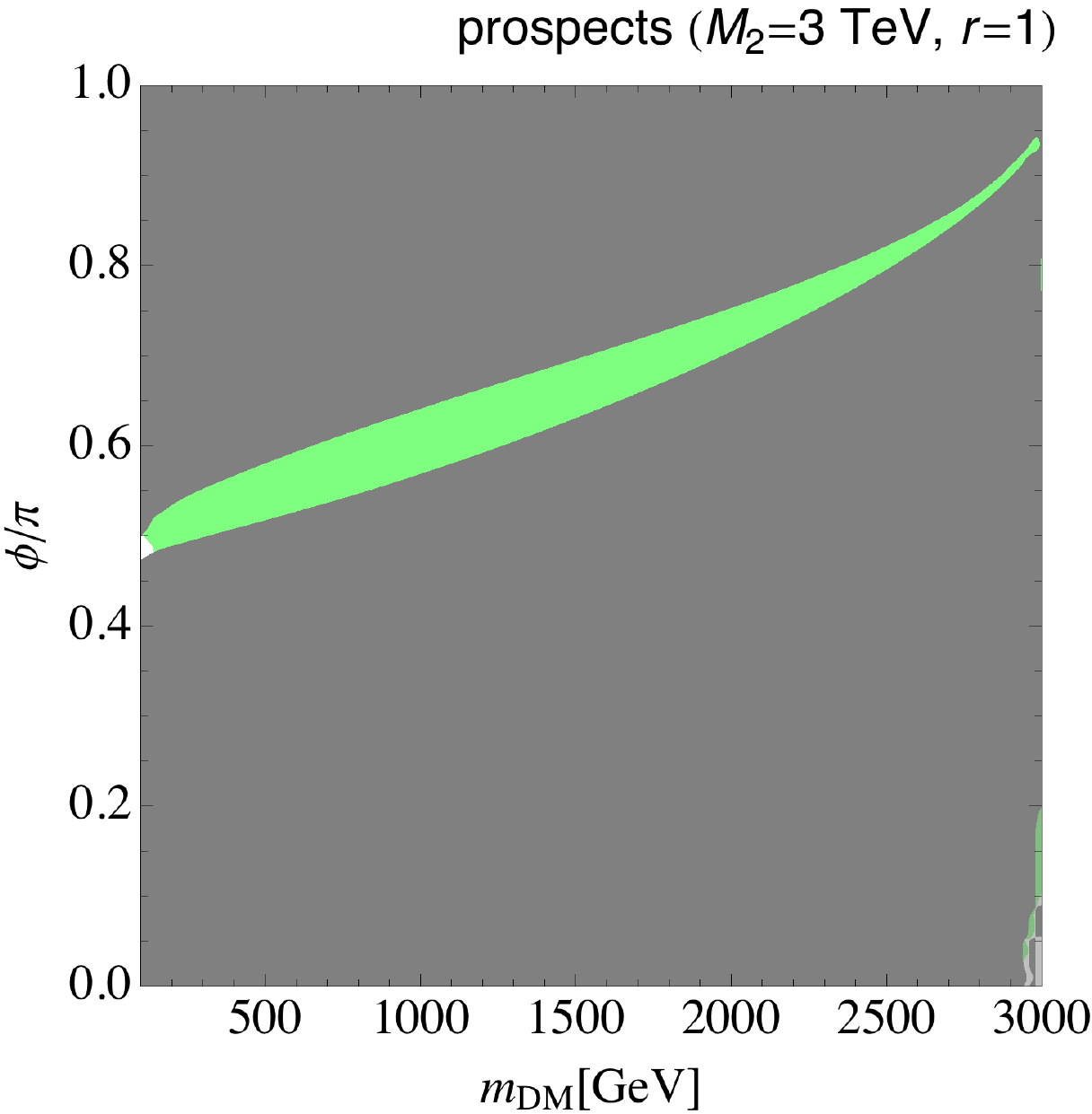}
\caption{ 
Current situation (upper panels) and prospects (lower panels)of the singlet-doublet model
for $M_2 = 3$~TeV and $r=1$.
The color notation is the same as Fig.~\ref{fig:result_M2=1TeV}.
}
\label{fig:result_M2=3TeV}
\end{figure}
Since the eEDM is proportional to $|M_2|^{-3}$ for $|M_2| \gg |M_1|, v$,
we expect that the constraints from the ACME experiment get milder for larger $M_2$.
We choose $M_2=3$~TeV in Fig.~\ref{fig:result_M2=3TeV} and in the lower panels in
Fig.~\ref{fig:result_cutoff}.
We find that the constraint from the eEDM measurement is weak compared with the case for $M_2=1$~TeV, as expected.
The value of the eEDM for $80 \lesssim m_{\text{DM}} \lesssim 100$~GeV 
is below the prospect~\cite{1208.4507,Kawall:2011zz}.
However, the stability bound is worse compared to the case for $M_2=1$~TeV as
can be seen from Fig.~\ref{fig:result_cutoff}.
In the large region, the cutoff scale is lower than 500~TeV for $M_2 = 3$~TeV, namely the uncertainty by ignoring the higher
dimensional operator is not negligible.
In particular,
50~TeV $<\Lambda<$ 500~TeV for $80 \lesssim m_{\text{DM}} \lesssim 100$~GeV, 
and thus the uncertainty is at most 10\%. 
The theoretical prediction in this region highly depends on how we modify the model
to push up the cutoff scale.

We conclude that the heavy dark matter region is already excluded by the
measurement of the eEDM. Most of the region except for the Higgs funnel region
is also excluded if experiments observe null results for the eEDM in future, or we
have to accept the ${\cal O}(10)$\% theoretical uncertainty in the calculation 
of the thermal relic abundance.

\begin{figure}[tb]
\includegraphics[width=0.45\hsize]{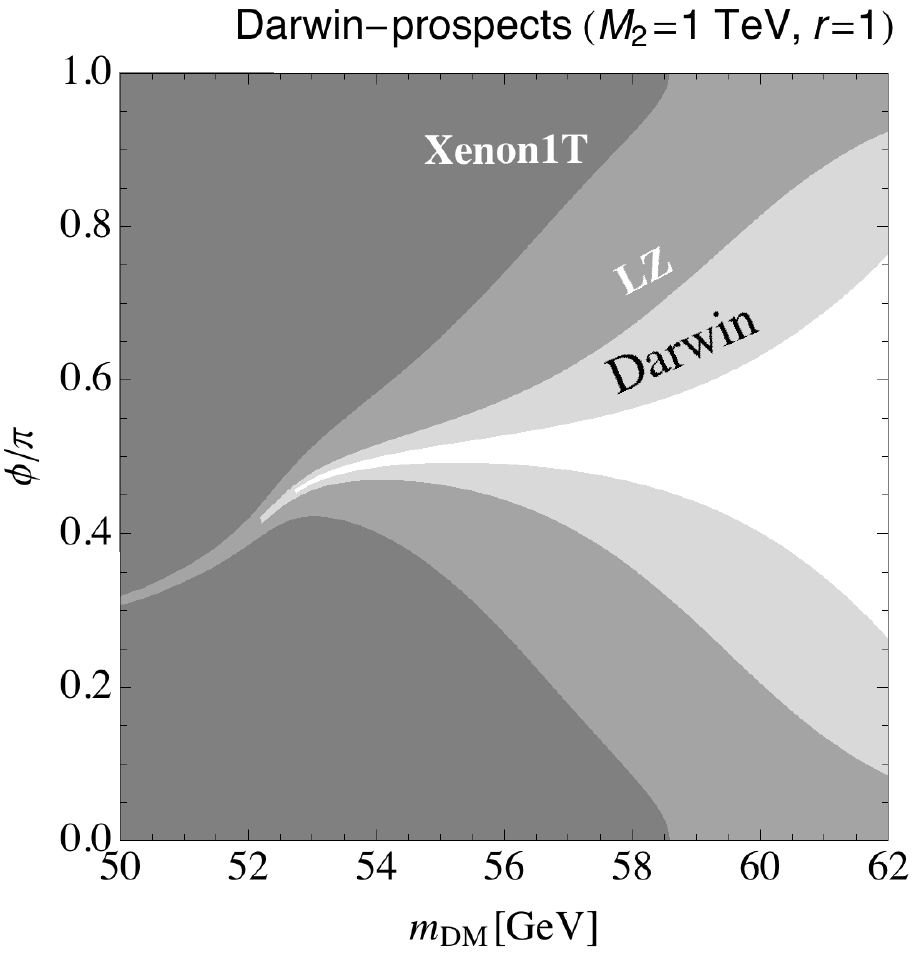}
\caption{The current and prospects of the direct detection experiments on the Higgs funnel region.}
\label{fig:darwin}
\end{figure}

We investigate how to test the model on the Higgs funnel region.
First, we focus on the progress of DM direct detection experiments.
The Darwin project is an ultimate experiment for the DM direct detection experiment~\cite{1606.07001}.
In Fig.~\ref{fig:darwin}, we show the prospect of the LZ and the Darwin in the Higgs funnel region.
The spin-independent cross section is highly suppressed in the Higgs funnel region.
Nevertheless, we find that the $CP$-conserving region is excluded by the LZ and Darwin.
On the other hand, 
the large $CP$-violating region cannot be tested by the direct detection experiment,
and thus we need other observables to study the Higgs funnel region completely.

\begin{figure}[tb]
\includegraphics[width=0.45\hsize]{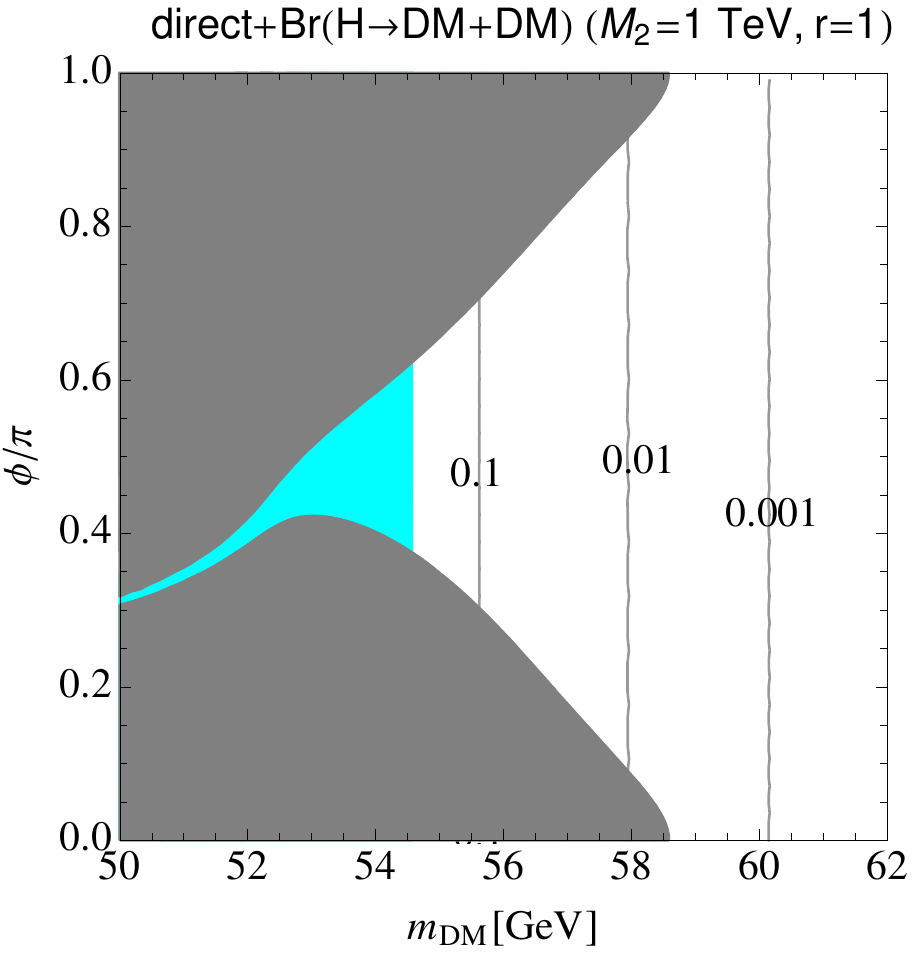}
\includegraphics[width=0.45\hsize]{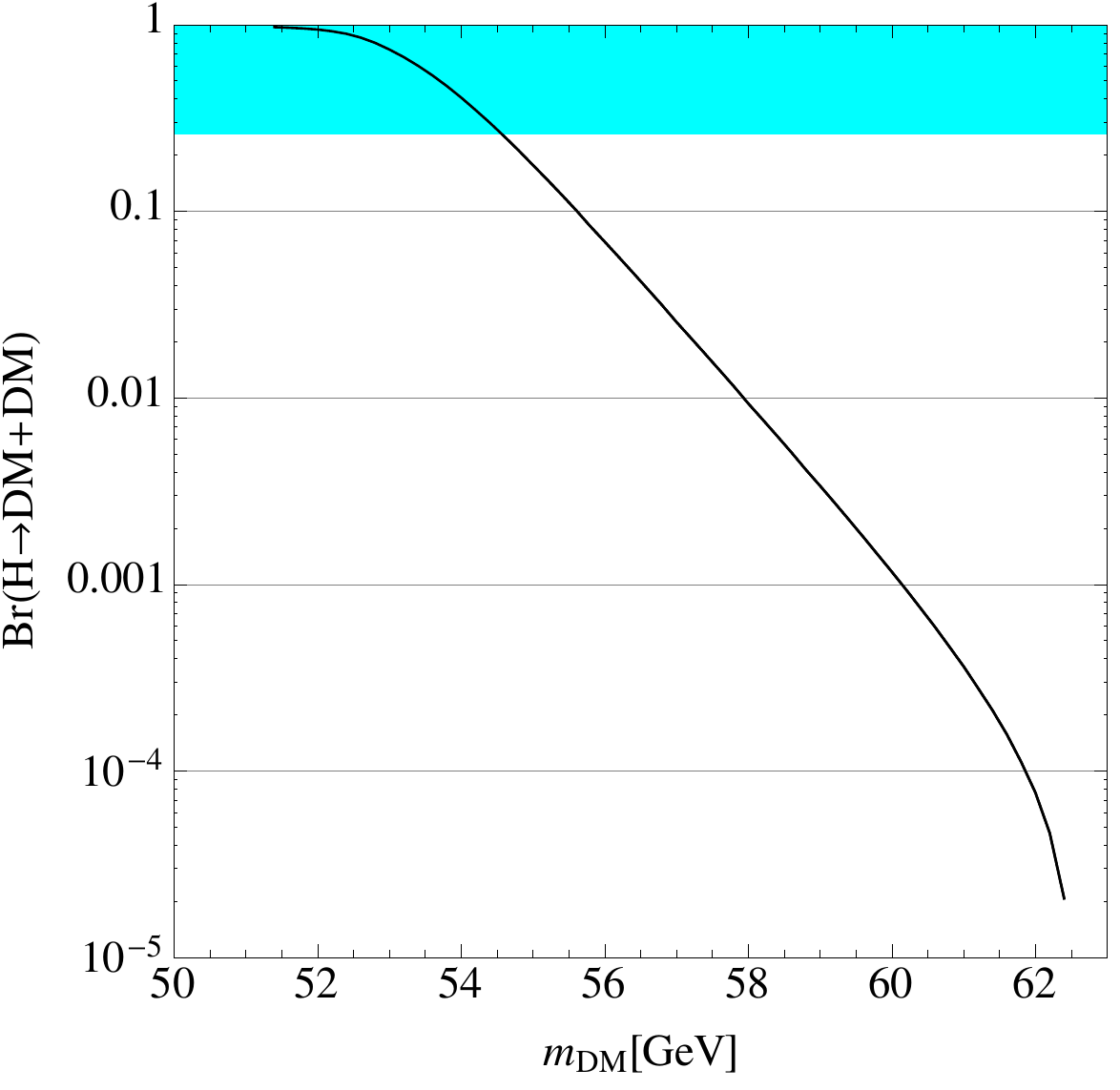}
\caption{ 
The branching ratio of the Higgs boson decaying to the DM pairs for $M_2 = 1$~TeV and $r = 1$.
The gray and cyan shaded regions are excluded by the XENON1T experiment~\cite{1805.12562}, 
and the LHC experiment~\cite{ATLAS-CONF-2018-054, 1809.05937}, respectively.
}
\label{fig:invisible}
\end{figure}

Second, we investigate the Higgs invisible decay.
The left panel of Fig.~\ref{fig:invisible} is a magnification of Fig.~\ref{fig:result_M2=1TeV} with 
the branching ratio of the Higgs decaying into a DM pair.
The right panel is the branching ratio for $\phi = 0.5 \pi$.
If the dark matter mass is very close to a half of the Higgs mass, 
the branching ratio becomes small due to the phase space suppression.
The prospects at future collider experiments on the upper bound on the invisible branching fraction are,
for example,
4.8~\% at the ILC ($\sqrt{s} = 250$ GeV, ${\cal L} = 250$ fb$^{-1}$) \cite{Asner:2013psa},
2.8~\% at the HL-LHC (${\cal L} = 3$ ab$^{-1}$) \cite{CMS:2016rfr},
and 0.5~\% at the FCC-ee ($\sqrt{s} = 240$ GeV, ${\cal L}=10$ ab$^{-1}$) \cite{Gomez-Ceballos:2013zzn}.
Therefore, the region $m_{\text{DM}} \lesssim 58$ GeV can be probed at the future collider experiments.

\begin{figure}[tb]
\includegraphics[width=0.45\hsize]{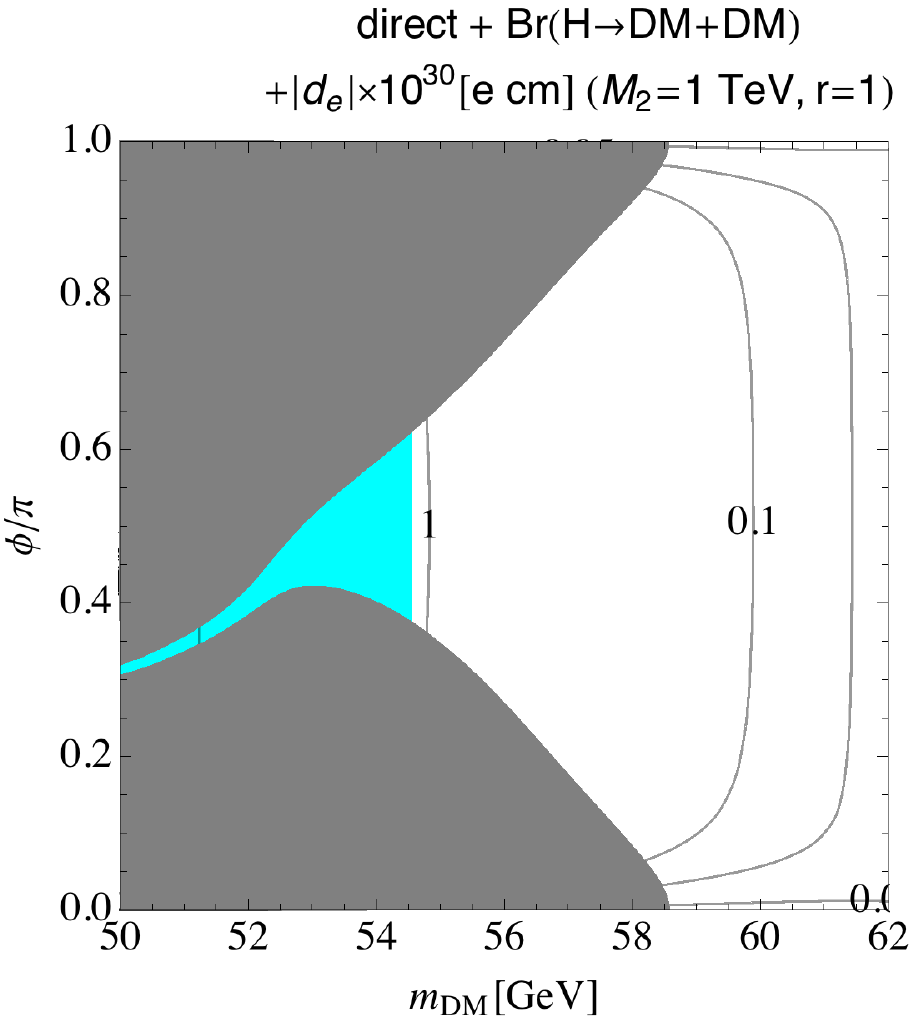}
\includegraphics[width=0.45\hsize]{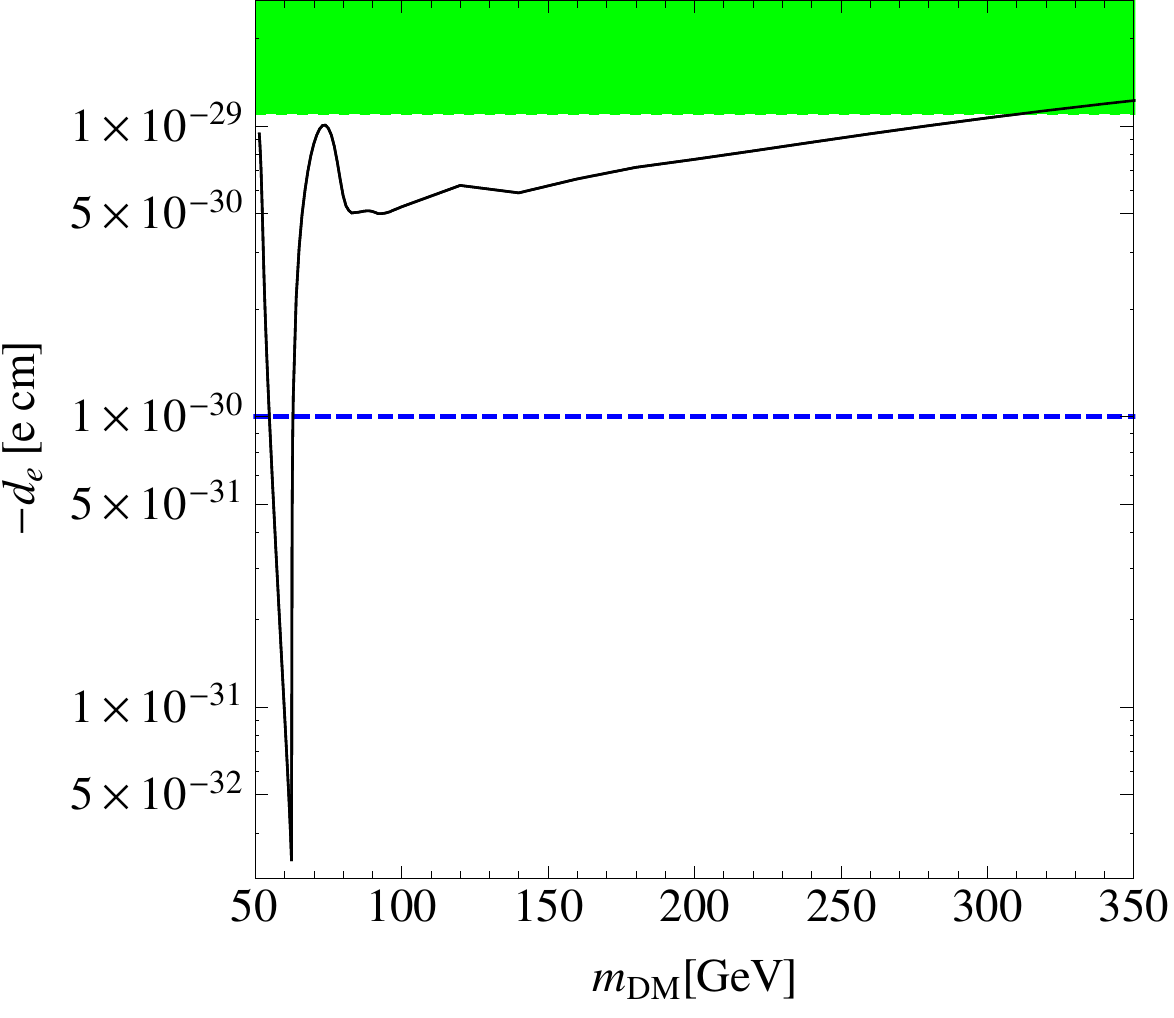}
\caption{ 
The eEDM for $M_2 = 1$~TeV, $r =1$.
The gray and cyan shaded regions are excluded by the XENON1T experiment~\cite{1805.12562}, 
and the LHC experiment~\cite{ATLAS-CONF-2018-054, 1809.05937}, respectively.
The green region in the right panel is excluded by the ACME experiment~\cite{Andreev:2018ayy}
The blue dashed line is the prospects by~\cite{1208.4507,Kawall:2011zz}.
We take $\phi = 0.5\pi$ in the right panel.
}
\label{fig:eEDM}
\end{figure}

Finally, let us discuss the eEDM.  Figure~\ref{fig:eEDM} shows eEDM in the Higgs funnel region.
We find that the eEDM 
changes rapidly for $0 < \pi < 0.1\pi$ and $0.9\pi < \phi < \pi$.
If the dark matter direct detection experiments observe null results in future,
the $CP$ invariance must be violated in the dark sector as shown in Fig.~\ref{fig:darwin}.
Consequently, the value of the eEDM must be nonzero if the DM-nucleon scattering cross section
is small enough to avoid the search by the direct detection experiments.
Assuming the null results by the direct detection experiments, 
we find that the minimum value of the eEDM is $d_e = 3 \times 10^{-32}~e$ cm 
for $|M_2|=1$~TeV and $r=1$
as can be seen from the right panel in Fig.~\ref{fig:eEDM}.
Therefore, we have a chance to test this model 
by the eEDM measurements  
in case that experiments reach to $d_e = {\cal O}(10^{-32})~e$ cm in future 
even if $\sigma_{\text{SI}}$ is very small at the Higgs funnel region. 
This is a different feature of this model from other Higgs portal DM models, which
avoid any constraints from the experiments and are not testable in the Higgs funnel region.

\section{Conclusion}
\label{sec:summary}
In this paper, we have investigated the singlet-doublet dark matter model with the $CP$ phase
as a UV complete model that predicts the pseudoscalar interaction. 
Although the scalar sector of the model is the same as the SM and does not contain
pseudoscalar fields, the model predicts the pseudoscalar interaction of the DM with the
SM Higgs boson because of the $CP$ violation in the dark sector.
As a result, the model evades the constraint from the DM direct detection experiments.
The DM annihilation processes into the SM particles with the pseudoscalar interaction
are $s$ wave, and its cross section at low temperature is not suppressed by powers of velocity.
As a result, we can expect the $\gamma$-ray emission from dSphs due to the annihilation of DM.
Besides, the $CP$ violation induces EDMs, which are severely constrained from experiments.
Consequently, some of the large $CP$ phase regions are already excluded, 
though the large $CP$ phase is helpful to avoid the constraint from the direct detection experiments.
Therefore, it is essential to investigate the correlation between 
the direct detection experiments and other observables that are sensitive to $CP$ violation.

We have calculated the $\gamma$-ray flux and compared with the result of the Fermi-LAT experiment
that gives the constraint of the $\gamma$-ray flux from dSphs.
We have used 19 dSphs whose J-factors were measured kinematically and listed in the sixth column in Table 1 in Ref.~\cite{Fermi-LAT:2016uux}. 
In this case, we find that the model is excluded for $m_h/2 <  m_{\text{DM}} < 74$~GeV.

We also have investigated the stability of the Higgs potential.
The new fermion fields in the dark sector give negative contributions to 
the beta function of the Higgs quartic coupling,
and thus the stability of the potential becomes worse than the SM case.
We have used the beta functions at the two-loop level with the threshold correction at the one-loop level,
and calculated the scale where the Higgs quartic coupling becomes 0, $\lambda(\Lambda)=0$.
The analysis of the effective theory for the fermionic DM models implies that 
the dimension-5 operators with the cutoff scale $\Lambda \lesssim 500$~TeV affect the calculation of the
DM annihilation cross section more than ${\cal O}(1)$\% level.  
Note that the DM energy density is measured with 1\% accuracy.
Therefore, our calculation is reliable for $\Lambda > 500$~TeV.
With this theoretical constraint, we find that 
our results do not receive the uncertainty from higher dimensional operators for
$m_{\text{DM}} < 1$~TeV and $|M_2| < 3$~TeV.
Although the model predicts a viable DM candidate without theoretical uncertainty for $\Lambda > 500$~TeV,
the Higgs potential is expected to be unstable in the large region of the 
parameter space because $\Lambda \ll \Lambda_{\text{SM}}$ in most of the
parameter space, where $\Lambda_{\text{SM}} \simeq 10^{10}$~GeV.
Therefore, an additional extension of the model is required to make the model
valid at a much higher energy scale such as the Planck scale.

We focused on the electron EDM.  
We found the region where $m_\text{DM} > 300 (2000)$~GeV is excluded for $|M_2|=1 (3)$~TeV.
Since the eEDM is proportional to $|M_2|^{-3}$ for $|M_2| \gg |M_1|, v$,
we can avoid the constraint from the eEDM with large $|M_2|$. 
However, the cutoff scale estimated from the stability bound of the Higgs potential
becomes lower for the large $|M_2|$ region. 
Therefore, we conclude that the heavy DM region is already excluded by the
measurement of the eEDM. 
We also found that most of the region except for the Higgs funnel region
is also excluded if experiments observe null results for $|d_e| \geq 10^{-30}~e$ cm in future, 
or we have to accept the ${\cal O}(10)$\% theoretical uncertainty in the calculation 
of the thermal relic abundance.

Finally, we studied the Higgs funnel region where 
the DM-nucleon scattering cross section and the annihilation cross section of DM
are highly suppressed. 
We find that the $CP$-conserving region can be covered by the LZ and Darwin experiments.
For the $CP$-violating region, we cannot expect the direct and indirect detection signals,
and thus other observables are required to test the model.
The searches for the Higgs invisible decay can probe for the dark matter
for $m_{\text{DM}} \lesssim 58$~GeV. Moreover, the eEDM is larger than $3 \times 10^{-32}~e$ cm
for $|M_2| = 1$~TeV, $r = 1$, and  $0.1 \pi < \phi < 0.9 \pi$.
If experiments for the eEDM search reach to ${\cal O}(10^{-32})~e$ cm, 
the Higgs funnel region can be probed by the combination of 
the direct detection, the searches for the Higgs invisible decay, and the eEDM.

\section*{Acknowledgments}
This work was supported by JSPS KAKENHI Grant Number 16K17715 [T.A.].

\appendix
\section{The threshold correction to the Higgs quartic coupling $\lambda$}\label{app:threshold}

We determine $\lambda$ from the on-shell Higgs mass given by
\begin{align}
 m_h^2 = m^2(\mu) + 3 \lambda(\mu) v^2(\mu) + (\text{loop corrections}).
\end{align}
If the theory changes at $\mu = M$, then 
\begin{align}
 m^2(M-0) + 3 \lambda(M-0) v^2(M-0) 
=&
 m^2(M+0) + 3 \lambda(M+0) v^2(M+0) + \Sigma^{\text{BSM}},
\end{align}
where $\Sigma^{\text{BSM}}$ is loop corrections to the Higgs mass from the new physics sector.
From this equation, we can find the threshold correction as follows,
\begin{align}
 \lambda(M+0) - \lambda(M-0)
=&
\lambda(M-0)
\left(
\frac{v^2(M-0)}{v^2(M+0)}-1
\right)
+
\frac{ m^2(M-0) -m^2(M+0) - \Sigma^{\text{BSM}}}{3 v^2(M+0)}
\nonumber\\
\simeq&
\lambda
\left(
\frac{-\Delta v^2}{v^2}
\right)
+
\frac{ - \Delta m^2 - \Sigma^{\text{BSM}}}{3 v^2(M+0)}
,
\end{align}
where
\begin{align}
 \Delta v^2 =& v^2(M+0) - v^2(M-0),\\
 \Delta m^2 =& m^2(M+0) - m^2(M-0).
\end{align}

$\Delta v^2$ is calculated from the Fermi constant that is determined from 
$\mu \to e \bar{\nu}_e \nu_\mu$.
We find
\begin{align}
 \Delta v^2
=&
 - \frac{4}{g^2} \Sigma_{WW}^{BSM}(0)
 - 2 v {\cal T},
\end{align}
where
\begin{align}
\Sigma^{BSM}_{WW}(0)
=&
- \frac{2}{(4\pi)^2}
\Biggl\{
\left( |c_{\chi_j}^L|^2 + |c_{\chi_j}^R|^2 \right)
\left( -2 B_{00}(0, m_{\chi_j}^2, m_{\chi^{\pm}}^2) + \frac{ m_{\chi_j}^2 + m_{\chi^{\pm}}^2}{2}
\right)
\nonumber\\
&
\qquad \qquad \quad
+
m_{\chi_j} m_{\chi^{\pm}} 
\left( (c_{\chi_j}^L)^* (c_{\chi_j}^R|^2) + (c_{\chi_j}^L) (c_{\chi_j}^R|^2)^* \right)
B_{0}(0, m_{\chi_j}^2, m_{\chi^{\pm}}^2)
\Biggr\}
,\\
 {\cal T}
=&
- \frac{1}{(4\pi)^2}
\frac{1}{m_h^2}
\sum_{j=1}^3 2 g_{\chi_j \chi_j h} m_{\chi_j} A_0(m_{\chi_j}^2).
\label{eq:tad}
\end{align}
We regularize the loop calculation by the $\overline{\text{MS}}$-scheme, 
and thus we have to keep calculate tadpole diagrams given in Eq.~\eqref{eq:tad}.
The loop functions ($A_0$, $B_0$, and $B_{00}$) are already regularized by the $\overline{\text{MS}}$-scheme,
and their definitions are given in \texttt{LoopTools}~\cite{hep-ph/9807565}. 
The couplings are given in Ref.~\cite{1702.07236}

$\Sigma^{\text{BSM}}$ is calculated from the loop correction to the Higgs mass from the new physics sector.
\begin{align}
 \Sigma^{\text{BSM}}
=&
 \Sigma_{\text{1PI}} + 6 \lambda v {\cal T},
\end{align}
where
\begin{align}
 \Sigma_{\text{1PI}}
=&
 \frac{2}{(4\pi)^2}
\Biggl\{
\sum_{j=1}^3
g_{\chi_j \chi_j h}^2
\left(
A_0(m_{\chi_j}^2)
+
\frac{4 m_{\chi_j}^2 - m_h^2}{2} B_0(m_h^2, m_{\chi_j}^2, m_{\chi_j}^2)
\right)
\nonumber\\
&
\qquad \quad
+
\sum_{j=1}^{3}
(g_{\chi_j \chi_j h}^A)^2
\left(
A_0(m_{\chi_j}^2)
+
\frac{ - m_h^2}{2} B_0(m_h^2, m_{\chi_j}^2, m_{\chi_j}^2)
\right)
\nonumber\\
&
\qquad \quad 
+
2 \sum_{j=1}^3 \sum_{k>j}^3
g_{\chi_j \chi_k h}^2
\left(
\frac{A_0(m_{\chi_j}^2) + A_0(m_{\chi_k}^2)}{2}
+
\frac{(m_{\chi_j} + m_{\chi_k})^2  - m_h^2}{2} B_0(m_h^2, m_{\chi_j}^2, m_{\chi_k}^2)
\right)
\nonumber\\
&
\qquad \quad 
+
2\sum_{j=1}^3 \sum_{k>j}^3
(g_{\chi_j \chi_k h}^A)^2
\left(
\frac{A_0(m_{\chi_j}^2) + A_0(m_{\chi_k}^2)}{2}
+
\frac{(m_{\chi_j} - m_{\chi_k})^2  - m_h^2}{2} B_0(m_h^2, m_{\chi_j}^2, m_{\chi_k}^2)
\right)
\Biggr\}
.
\end{align}

We choose $m^2$ so as to satisfy that the effective potential has the electroweak vacuum at the one-loop level.
The effective potential at the one-loop level is given by
\begin{align}
 V_{eff.}
=&
\frac{m^2}{2} \varphi^2 + \frac{\lambda}{4} \varphi^4
\nonumber\\
&
- \frac{1}{64\pi^2} 
\left(
3 (m^2 + \lambda \varphi^2)^2 
\left(
\frac{3}{2} + \ln \frac{\mu^2}{m^2 + \lambda \varphi^2}
\right)
\right)
\nonumber\\
&
- \frac{1}{64\pi^2} 
\left(
(m^2 + 3 \lambda \varphi^2)^2 
\left(
\frac{3}{2} + \ln \frac{\mu^2}{m^2 + 3 \lambda \varphi^2}
\right)
\right)
\nonumber\\
&
- \frac{3}{64\pi^2} 
\left(
2 \left(\frac{g^2 \varphi^2}{4}\right)^2 
\left(
\frac{5}{6} + \ln \frac{\mu^2}{\frac{g^2 \varphi^2}{4}}
\right)
\right)
\nonumber\\
&
- \frac{3}{64\pi^2} 
\left(
\left(\frac{(g^2 + g'^2)\varphi^2}{4}\right)^2 
\left(
\frac{5}{6} + \ln \frac{\mu^2}{\frac{(g^2 + g'^2)\varphi^2}{4}}
\right)
\right)
\nonumber\\
&
+ \frac{12}{64\pi^2} 
\left(
\left(\frac{ y_t^2 \varphi^2}{2}\right)^2 
\left(
\frac{3}{2} + \ln \frac{\mu^2}{\frac{y_t^2 \varphi^2}{2}}
\right)
\right)
\nonumber\\
&
+ \frac{2}{64\pi^2} 
\sum_{j=1}^3
\left(
m_{\chi_j}^4
\left(
 \frac{3}{2} + \ln \frac{\mu^2}{m_{\chi_j}^2}
\right)
\right)
.
\end{align}
We choose $m^2$ so as to satisfy $V_{eff.}'(v) =0$ at the one-loop level.
\begin{align}
m^2
=&
- \lambda v^2
+
\frac{1}{(4\pi)^2}
\Bigg\{
 6 \lambda^2 v^2 \left( 1 + \ln \frac{\mu^2}{2 \lambda v^2} \right)
\nonumber\\
 & \qquad \qquad \qquad +
 6 v^2 \left(\frac{g^2}{4}\right)^2 \left( \frac{1}{3} + \ln \frac{\mu^2}{\frac{g^2 v^2}{4}} \right)
+3 v^2 \left(\frac{g^2+g'^2}{4}\right)^2 \left( \frac{1}{3} + \ln \frac{\mu^2}{\frac{(g^2 +g'^2)v^2}{4}} \right)
\nonumber\\
 & \qquad \qquad \qquad
 -12 v^2 \left(\frac{y_t^2}{2}\right)^2 \left( 1 + \ln \frac{\mu^2}{\frac{y_t^2 v^2}{2}} \right)
\nonumber\\
 & \qquad \qquad \qquad
- 2\frac{1}{v} \sum_{j} g_{\chi_j \chi_j h} m_{\chi_j}^3 \left(1 + \ln \frac{\mu^2}{m_{\chi_j}^2} \right)
\Biggr\}.
\label{eq:m2}
\end{align}
From Eq.~\eqref{eq:m2}, we find
\begin{align}
 \Delta m^2
=&
 \frac{1}{(4\pi)^2}
\left[
- 2\frac{1}{v} \sum_{j} g_{\chi_j \chi_j h} m_{\chi_j}^3 \left(1 + \ln \frac{\mu^2}{m_{\chi_j}^2} \right)
\right].
\end{align}

Finally, we find
\begin{align}
 \lambda(M+0) - \lambda(M-0)
\simeq&
- \frac{1}{3 v^2}
\left(
\Sigma_{\text{1PI}} + 2 \lambda v {\cal T}
\right)
+
\frac{\lambda}{v^2}
\frac{4}{g^2} \Sigma_{WW}^{\text{BSM}}(0)
.
\end{align}


\end{document}